\documentclass[graybox]{svmult}

\usepackage[utf8]{inputenc}
\usepackage[T1]{fontenc}
\usepackage{geometry}
\usepackage{xcolor}
\usepackage{prettyref}
\usepackage{float}
\usepackage{units}
\usepackage{mathrsfs}
\usepackage{dsfont}
\usepackage{amsmath}
\usepackage{bm} 
\usepackage{amssymb}
\usepackage{mathtools}
\usepackage{lmodern}
\usepackage{circuitikz}
\usepackage{qcircuit}
\usepackage{dsfont}
\usepackage{xfrac}
\usepackage{wasysym}
\usepackage{cancel}
\usepackage{stackrel}
\usepackage{graphicx}
\usepackage{esint}
\usepackage{comment}
\usepackage{enumitem}
\usepackage{type1cm} 
\usepackage{makeidx} 
\usepackage{graphicx} 
\usepackage{multicol} 
\usepackage[bottom]{footmisc}

\usepackage{newtxtext} %

\usepackage{esvect}
\usepackage{physics}
\usepackage[backref=page]{hyperref}
\hypersetup{
    colorlinks=true,
    linkcolor=blue,
    filecolor=magenta,      
    urlcolor=blue,
    citecolor=red,
    pdftitle={Entanglement review},
    }

\usepackage{pdflscape}
\usepackage{babel}

\makeatletter

\newcommand{\Id}{1}
\newcommand{\X}[1]{\textsf{X}_{#1}}
\newcommand{\Y}[1]{\textsf{Y}_{#1}}
\newcommand{\Z}[1]{\textsf{Z}_{#1}}
\newcommand{\T}{\textsf{T}}
\renewcommand{\S}{\textsf{S}}

\newcommand{\CNOT}{\textsf{CNOT}}

\newcommand{\CPHASE}{\textsf{CPHASE}}
\newcommand{\CZ}{\textsf{CZ}}

\begin{document}

\newpage

\title*{Josephson junctions, superconducting circuits, and qubit for quantum technologies}

\author{Roberta Citro and Claudio Guarcello and Sergio Pagano}
\institute{R. Citro\at Department of Physics ``E.R. Caianiello'', University of Salerno, Fisciano, Italy \email{rocitro@unisa.it}
\and C. Guarcello \at Department of Physics ``E.R. Caianiello'', University of Salerno, Fisciano, Italy \email{cguarcello@unisa.it}
\and S. Pagano \at Department of Physics ``E.R. Caianiello'', University of Salerno, Fisciano, Italy \email{spagano@unisa.it}}

\maketitle




\begin{abstract}\\
In the realm of physics, a pivotal moment occurred six decades ago when Brian Josephson made a groundbreaking prediction, setting in motion a series of events that would eventually earn him the prestigious Nobel Prize eleven years later. This prediction centered around what is now known as the Josephson effect, a phenomenon with far-reaching implications. At the heart of this effect lies the Josephson junction (JJ), a device that has become a linchpin in various scientific applications.
This chapter delves into the foundational principles of the Josephson effect and the remarkable properties of JJs. From their role in metrology to their application in radiation detectors, these junctions have ushered in a new era of electronics. Exploiting the unique features of superconductive devices, such as high speed, low dissipation, and dispersion, JJs find today practical implementation in the development of superconductive qubits and nanotechnology applications.
\end{abstract}

\renewcommand{\thefootnote}{\fnsymbol{footnote}}
\footnotetext[1]{This is a preprint of the following chapter: Roberta Citro, Claudio Guarcello, and Sergio Pagano ''\emph{Josephson junctions, superconducting circuits, and qubit for quantum technologies}", to be published in \emph{New Trends and Platforms for Quantum Technologies} edited by Ramon Aguado, Roberta Citro, Maciej Lewenstein and Michael Stern, 2024, Springer reproduced with permission of Springer Nature Switzerland AG. The final authenticated version is available online at: \href{http://dx.doi.org/10.1007/978-3-031-55657-9}{http://dx.doi.org/10.1007/978-3-031-55657-9}}
\renewcommand{\thefootnote}{\arabic{footnote}}

\section{The Josephson effect}

The Josephson effect is a quantum phenomenon observed in superconducting systems. In its simplest form, the Josephson effect involves the flow of supercurrent (a current without resistance) between two superconductors separated by a thin insulating barrier, typically a JJ. The key feature of the Josephson effect is the phase coherence of the wave functions of the superconducting electrons on either side of the barrier, as we will see in this Chapter.
There are two main types of Josephson effects: a) {\it dc Josephson effect} when no voltage is applied across the JJ, resulting in a steady supercurrent flow through the junction. The current is directly proportional to the sine of the phase difference between the wave functions of the superconducting electrons on either side of the barrier; b) {\it ac Josephson effect} when a fixed voltage is applied across the JJ, leading to an alternating supercurrent that oscillates at the same frequency as the applied voltage. This phenomenon is used in various applications such as superconducting quantum interference devices (SQUIDs), Josephson voltage standards, and superconducting qubits for quantum computing.\\
The Josephson effect has significant practical applications in areas such as metrology, quantum computing, and high-speed digital electronics. It also provides insights into the fundamental properties of superconductors and quantum mechanics.

\subsection{First Josephson equation}

When thinking about the derivation of the Josephson equations, one usually refers to Feynman's formulation~\cite{Feynman1965}, which is based on a ``two-level system'' picture. The interested reader can find a comprehensive overview of this method in the Barone and Patern\'o's book~\cite{Barone1982}. Here, we follow a different line of thinking, proposing an approach due to L.D. Landau and E.M. Lifschitz~\cite{Landau1980} and later discussed in Ref.~\cite{Gross2016}. 

Let's start with some considerations. First, it is reasonable to assume that the supercurrent depends on the density of Cooper pairs in the superconductors forming the junction, $\left|\Psi_1 \right|^2 = n_{s,1}^\star $ and $\left|\Psi_2 \right|^2 = n_{s,2}^\star $. 
Furthermore, since the coupling between the two superconductors is ``weak'', we can also assume that the supercurrent density between the two superconducting electrodes does not change $\left|\Psi \right|^2$. On the other side, it is reasonable to expect that the supercurrent density depends on the phases of the wave functions. 
In a bulk superconductor, the supercurrent density is proportional to the gauge-invariant phase gradient $\gamma(\textbf{r},t)$ according to
\begin{equation}\label{Chapter1:eq:J_s}
 J_s(\textbf{r},t)=\frac{q^\star n_s^\star \hbar}{m^\star}\gamma(\textbf{r},t) \qquad\text{where}\qquad \gamma(\textbf{r},t)=\nabla \theta - \frac{2\pi}{\Phi_0}\textbf{A}(\textbf{r},t).
\end{equation}
Here, $\Phi_0=h/(2e)$ is the flux quantum, $\hbar=h/(2\pi)$ is the reduced Planck constant, $\textbf{A}$ is the potential vector, $m^\star$ and $q^\star$ are the charge and the mass of superelectrons, respectively.

We simplify our discussion by introducing two key assumptions. Firstly, we posit that the current density is uniformly distributed, namely, the junction area is sufficiently small. Secondly, we consider two weakly connected superconductors and that the phase gradient in the superconducting electrodes undergoes negligible variation: this condition holds when the Cooper pairs density in the electrodes is greater than in the coupling region. Since the supercurrent density remains constant in the electrodes (owing to current conservation), the only relevant phase gradient is that in the interlayer region, in accordance with Eq.~\eqref{Chapter1:eq:J_s}. Thus, we can focus on the \emph{gauge-invariant phase difference}, given by
\begin{equation}\label{Chapter1:eq:phi1}
\varphi(\textbf{r},t)=\int_1^2\gamma(\textbf{r},t)=\int_1^2\left ( \nabla \theta(\textbf{r},t) - \frac{2\pi}{\Phi_0}\textbf{A}(\textbf{r},t) \right )d\textbf{l}= \theta_2(\textbf{r},t)-\theta_1(\textbf{r},t) - \frac{2\pi}{\Phi_0}\int_1^2\textbf{A}(\textbf{r},t)\, d\textbf{l} ,
\end{equation}
with the integration path along the flow direction of the current. One can reasonably assume that the supercurrent density is proportional to the phase difference, i.e, $J_s=J_s(\varphi)$. Moreover, since any $2\pi$-phase shift yields an identical wave function of the superconducting electrodes, we can expect even $J_s(\varphi)$ to be a $2\pi$-periodic function, i.e., $J_s(\varphi)=J_s(\varphi+2\pi\,n)$.
Finally, with no current applied $\varphi=0$, i.e. $\theta_1 = \theta_2$, and thus $J_s(0)=J_s(2\pi\,n)=0$.
Summing up, we can say that
\begin{equation}\label{Chapter1:eq:J_s2}
J_s(\varphi)=J_c \sin\varphi+\sum_{m=2}^\infty J_m\sin(m\varphi),
\end{equation}
with $J_c$ being the Josephson critical current density, which depends on the coupling between the superconductors. In general, time-invariance of the Josephson current requires that $J_s(\varphi)=-J_s(-\varphi)$, and this is why all cosine terms in the Fourier series forming $J_s(\varphi)$ are ignored. Equation~\eqref{Chapter1:eq:J_s2} is usually refereed to as the \emph{1$^{st}$ Josephson relation}, or even as the \emph{current-phase relation} (CPR). Often, only the first term of the expansion is relevant, i.e., 
\begin{equation}
J_s(\varphi)=J_c \sin\varphi,
\end{equation}
as it was derived by Josephson in his original article for an insulating barrier~\cite{Josephson1962,Josephson1974}.

So far we performed a 1D analysis by considering a uniform supercurrent density: however, the reasoning presented remains still valid if applied to each point $(y, z)$ of the junction area. In particular, we can assume $J_c=J_c(y, z)$ and $J_s$ flowing perpendicularity to the junction area for any given $y$ and $z$, since the current flows in the $x$-direction. Thus, the current density may vary with $y$ and $z$ and the CPR has to be generalized to
\begin{equation}\label{Chapter1:eq:J_s3}
J_s(y,z,t)=J_c(y,z) \sin\varphi(y,z,t).
\end{equation}

A tunnel JJ represents a sort of bottleneck for the current density in the superconducting channel, being the Josephson critical current, i.e., the maximum Josephson current that can flow without triggering a voltage state, far smaller than the usual depairing current in superconducting electrodes~\cite{Likharev1979}; from this aspect comes the term ``weak link", which serves to capture the idea that the JJ is a region where deviations from ideal superconducting behavior are most significant, impacting both the tunneling of Cooper pairs (Josephson effect) and the stability of superconductivity (depairing).

\begin{figure}[htbp!!]
\centering{}\includegraphics[width=0.5\columnwidth]{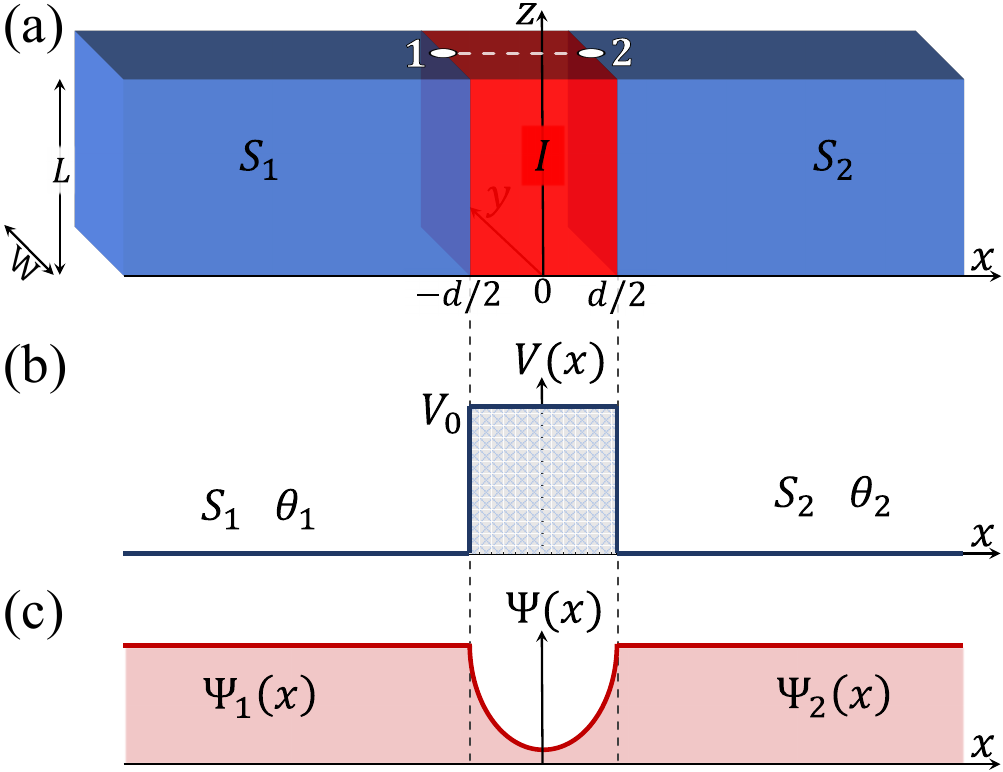}\caption{(a) Sketch of an SIS junction, formed by sandwiching two superconductors, $S_1$ and $S_2$, on an insulating layer, $I$, of thickness $d$. The junction area $A=LW$ lies in the $(y,z)$ plane, while the current flow through $A$ occurs along $x$. (b) Potential barrier $V(x)=V_0$ for $\left| x\right|\leq d/2$ and $0$ otherwise. (c) Sketching of the profile of a wave function $\Psi(x)$, see Eq~\eqref{Chapter1:eq:SchrodingerEqSolutions_1}.}
\label{Chapter1:fig:JosephsonEffect00}
\end{figure}

\subsection{Second Josephson equation}

To derive the link between the voltage drop and the Josephson phase, we start from the time derivative of Eq.~\eqref{Chapter1:eq:phi1},
\begin{equation}\label{Chapter1:eq:phi_t1}
 \frac{\partial \varphi}{\partial t}= \frac{\partial \theta_2}{\partial t} - \frac{\partial \theta_1}{\partial t}- \frac{2\pi}{\Phi_0} \frac{\partial }{\partial t}\int_1^2\textbf{A}\, d\textbf{l}\, .
\end{equation}
Then, we recall that in a bulk superconductor, with $n_{s}^\star=constant$, the following relation holds:
\begin{equation}\label{Chapter1:eq:theta_t1}
-\hbar \frac{\partial \theta}{\partial t}= \frac{1}{2n_s^{\star}}\Lambda J_s^2+q^{\star}\varphi.
\end{equation}
Inserting this expression into Eq.~\eqref{Chapter1:eq:phi_t1} and for a continuous supercurrent density, i.e., $J_s(2)=J_s(1)$, one obtains
\begin{equation}
 \frac{\partial \varphi}{\partial t}= \frac{2\pi}{\Phi_0} \int_1^2\left ( -\nabla \phi-\frac{\partial \textbf{A}}{\partial t}\right ) d\textbf{l}.
\end{equation}
Since the contribution in parentheses is nothing more than the electric field, this equation can be recast as
\begin{equation}
 \frac{\partial \varphi}{\partial t}= \frac{2\pi}{\Phi_0} \int_1^2\textbf{E}(\textbf{r},t) d\textbf{l},
\end{equation}
obtaining the so-called \emph{2$^{nd}$ Josephson relation}, or, alternatively, the \emph{voltage-phase relation}. In fact, the integral in this equation is the potential difference $V$, which turns out in the chemical potential difference, $\Delta\mu=\mu_2-\mu_1=eV$, between the two superconductors. 
In the case of a constant voltage drop applied to the junction, one simply obtain
\begin{equation}
 \frac{\partial \varphi}{\partial t}= \frac{2\pi}{\Phi_0} V,
\end{equation}
that is the Josephson phase grows linearly in time according to
\begin{equation}
\varphi(t)= \varphi_0+ \frac{2\pi}{\Phi_0} V\,t.
\end{equation}
In this case, the Josephson current oscillates in time at a specific frequency, $\nu/V=\Phi_0^{-1}\simeq483.6\;\text{MHz}/\mu\text{V}$. In other words, a JJ can work as a voltage-controlled oscillator capable of producing very high frequencies ($\sim500\;\text{GHz}$ at $1\;\text{mV}$). According to these equations, exciting a junction with a signal at a frequency $\nu$ would result in the emergence of constant voltage regions on its current-voltage (IV) curve at values $n\Phi_0\nu$. Shapiro experimentally validated this prediction in 1963~\cite{Shapiro1963}, and it is now recognized as the \emph{ac Josephson effect}. The practical implications of this phenomenon were promptly applied to metrology, as it establishes a connection between the volt and the second via a proportionality reliant solely on fundamental constants. Initially, this aspect contributed to a refined determination of the $h/e$ ratio. Presently, it serves as the foundation for voltage standards globally~\cite{Hamilton2000}.

\subsection{Estimation of the maximum Josephson current density}

This section discusses the determination, with general arguments, of the maximum value of the Josephson current density, i.e., the \emph{critical current density} $J_c$, in the case of a superconductor-insulator-superconductor (SIS) junction, see Fig.~\ref{Chapter1:fig:JosephsonEffect00}(a). In particular, we address this situation with the so-called \emph{wave-matching method}, i.e., solving the Schr\"{o}dinger problem in both the superconducting electrodes and in the insulating region and looking at the boundary conditions to determine the coefficients for matching the solutions in these regions. 

Looking at the superconducting electrodes, the supercurrent density is given by Eq.~\eqref{Chapter1:eq:J_s} at the positions $x=\pm d/2$. 
In the absence of electric/magnetic fields, from Eq.~\eqref{Chapter1:eq:theta_t1} we get
\begin{equation}
 \frac{\partial \theta}{\partial t}= -\frac{\Lambda J_s^2}{2n_s^{\star}\hbar}=-\frac{E_0}{\hbar},
\end{equation}
with $E_0=m^\star v_s^2/2$ being the kinetic energy of superelectrons. The time-dependent, macroscopic wave function can be expressed as $\Psi(\textbf{r},t)=\Psi(\textbf{r})\, \text{exp} \left [ -i(E_0/\hbar)t \right ]$,
where $\Psi(\textbf{r})$ indicates its time-independent amplitude. 

Let's consider now the insulating region. As depicted in Fig.~\ref{Chapter1:fig:JosephsonEffect00}(b), the potential is considered to be zero outside this area and equal to $V_0>E_0$ inside it. We assume only elastic tunneling processes, that is in which superelectrons conserve energy, so that the time evolution of $\Psi(\textbf{r},t)$ is the same outside and inside the barrier and it is therefore sufficient to regard the time-independent part of the problem. Thus, we can write down a time-independent Schr\"{o}dinger-like equation 
\begin{equation}\label{Chapter1:eq:SchrodingerEq_1}
-\frac{\hbar^2}{2m^\star}\nabla^2\Psi(\textbf{r})=(E_0-V_0)\Psi(\textbf{r}).\end{equation}
At this point, we need to additional assumptions: \emph{i}) a uniform tunneling barrier and \emph{ii}) a small junction area $A=LW$, so that the Josephson current density can be considered uniform within $A$ and Eq.~\eqref{Chapter1:eq:SchrodingerEq_1} reduces just to a one-dimensional problem in the $x$ direction, whose solution is
\begin{equation}\label{Chapter1:eq:SchrodingerEqSolutions_1}
\Psi(x)=A\cosh\left ( \kappa x \right )+ B \sinh\left ( \kappa x \right ),
\end{equation}
where the \emph{characteristic decay constant} $\kappa$ is defined as
\begin{equation}
\kappa=\sqrt{\frac{2m^\star}{\hbar^2}\left ( V_0-E_0 \right )}.
\end{equation}
Taking into account the boundary conditions,
\begin{equation}
\Psi\left ( -d/2 \right )=\sqrt{n_1^\star}e^{i\theta_1}\qquad\text{and}\qquad\Psi\left ( d/2 \right )=\sqrt{n_2^\star}e^{i\theta_2},
\end{equation}
the coefficients become
\begin{equation}\label{Chapter1:eq:SchrodingerEqCoefficients_1}
A=\frac{\sqrt{n_1^\star}e^{i\theta_1}+\sqrt{n_2^\star}e^{i\theta_2}}{2\cosh\left ( \kappa d/2 \right )}\qquad\text{and}\qquad B=\frac{\sqrt{n_1^\star}e^{i\theta_1}-\sqrt{n_2^\star}e^{i\theta_2}}{2\sinh\left ( \kappa d/2 \right )}.
\end{equation}
Finally, inserting the wave function expression given in Eq.~\eqref{Chapter1:eq:SchrodingerEqSolutions_1} into the supercurrent density equation yields
\begin{equation}
J_s=\frac{q^\star}{m^\star}\Re \left\{ \Psi^\star\left ( \frac{\hbar}{i}\nabla \right )\Psi\right\}=\frac{q^\star}{m^\star}\Im \left\{ A^\star B\right\}.
\end{equation}
Now, by considering the specific coefficients in Eq.~\eqref{Chapter1:eq:SchrodingerEqCoefficients_1}, one obtains the supercurrent density $J_s=J_c \sin (\theta_2-\theta_1)$, where
\begin{equation}\label{Chapter1:eq:Jc_1}
J_c=-2\frac{q^\star \kappa\hbar}{m^\star}\frac{\sqrt{n_1^\star\,n_2^\star}}{2\sinh\left (2 \kappa d \right )}.
\end{equation}

A $V_0$ of the order of a few eV gives a decay length, $1/\kappa$, less than a nanometer, so that $\kappa d\gg 1$, if the barrier thickness $d$ is just a few nanometers. Using the hyperbolic identity $2\sinh(x)\cosh(x)=\sinh(2x)$, one obtains $2\sinh(\kappa d/2)\cosh(\kappa d/2)=\sinh(\kappa d)$. Thus, in the limit $\kappa d\gg 1$, one can approximate $\sinh(\kappa d)\simeq \exp(\kappa d)/2$ in Eq.~\eqref{Chapter1:eq}, obtaining
\begin{equation}\label{Chapter1:eq:Jc_2}
J_c\simeq 2\frac{e\kappa\hbar}{m}\sqrt{n_1^\star,n_2^\star},e^{-\kappa d},
\end{equation}
where we replaced $q^\star=-2e$ and $m^\star=2m$. In other words, with simple assumptions we achieved Eq.~\eqref{Chapter1:eq:Jc_2} describing a supercurrent density that decays exponentially as the thickness of the insulating layer is increased.

\subsection{Anomalous Josephson effect}

The CPR represents the way to calculate most of junction properties. Anyway, only in a few cases the CPR reduces to the familiar sinusoidal form $I_s (\phi) = I_c \sin(\phi)$, which is however ordinarily used to study the dynamics and performance of devices based on conventional JJs.

There are several properties of the CPR that are rather general and depend neither on the junction's materials and geometry nor on the theoretical model used to describe the processes in the junction~\cite{Golubov2000}:
\begin{itemize}
    \item[1)] A change of phase of the order parameter of $2\pi$ in any of the electrodes is not accompanied by a change in their physical state. Consequently, this change must not influence the supercurrent across a junction, and $I_s (\varphi)$ should be a $2\pi$ periodic function, $I_s (\varphi) = I_s (\varphi + 2\pi)$;
    \item[2)] Changing the direction of supercurrent entails a change of the sign of $\varphi$; therefore $I_s (\varphi) = -I_s (-\varphi)$. Note that this condition is violated in superconductors with broken time-reversal symmetry (TRS), and this can lead to spontaneous currents;
    \item[3)] A dc supercurrent can flow only if there is a gradient of the order-parameter phase. Hence, in the absence of phase difference, there should be zero supercurrent, $I_s (2\pi n)=0$, with $n=0,\pm1,\pm2,\ldots$;
    \item[4)] It follows from 1) and 2) that the supercurrent should also be zero at $\varphi=\pi n$, that is $I_s (\pi n)=0$, for $n=0,\pm1,\pm2,\ldots$; therefore, it is sufficient to consider $I_s (\varphi)$ only in the interval $0<\varphi<\pi$.
\end{itemize}

Thus, $ I_s (\varphi) $ can be in general decomposed into a Fourier series, i.e.,
\begin{equation}\label{Chapter1:eqn:GeneralCPR}
I_s (\varphi) = \sum_{n \geq 1} [I_n \sin(n\varphi) + J_n \cos(n\varphi)].
\end{equation}
The $I_n$ term depends on the barrier transparency $D$ as a $D^n$ power-law and corresponds to the $n$-multiple Andreev reflection process, thus these terms become more relevant when increasing the barrier transparency.
For satisfying the condition 2), the term $ J_n \cos(n\varphi) $ has to be zero, that is, in other words, $J_n$ is present only if TRS is broken.

\begin{figure}[t!!]
\centering{}\includegraphics[width=0.85\columnwidth]{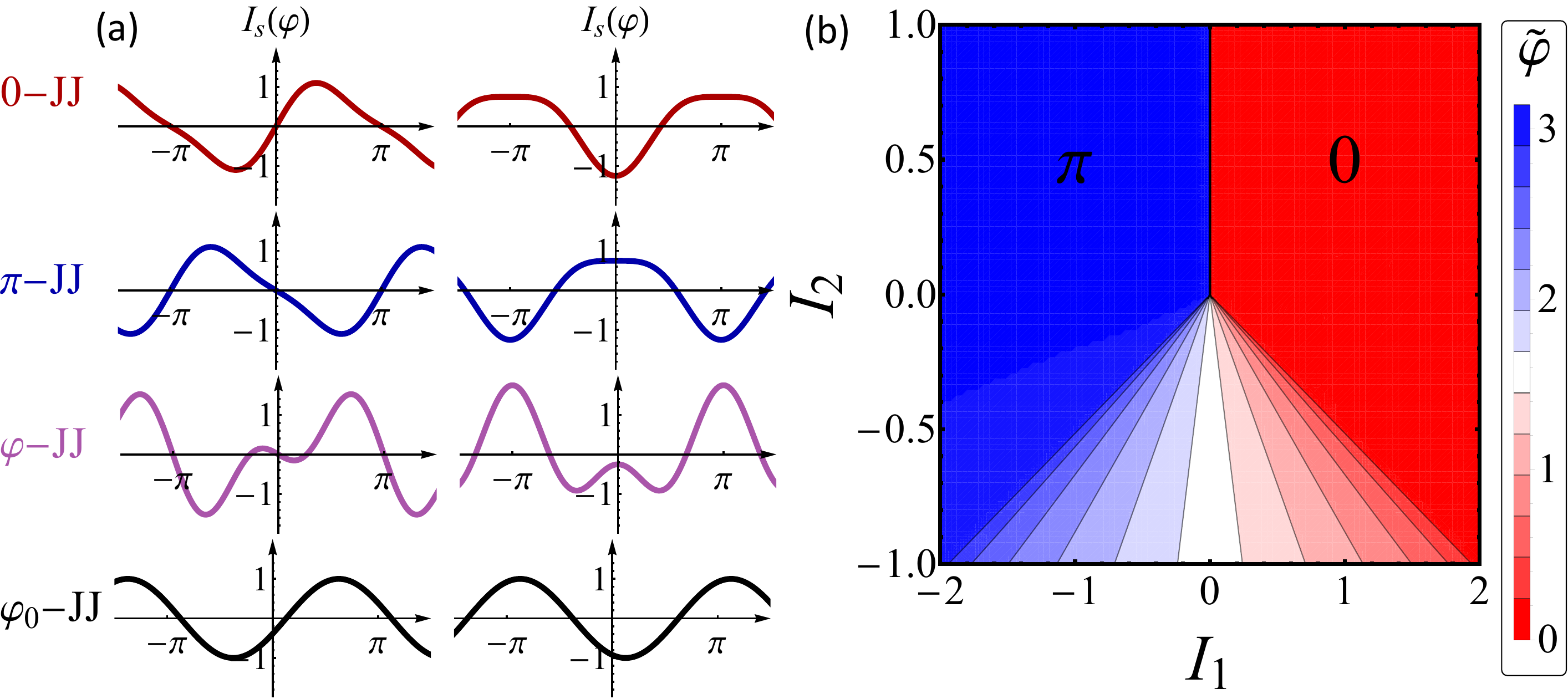}
\caption{(a) CPRs, $I_s(\varphi)$, and Josephson energies, $F_J(\varphi)$, in the case of a $0-$ (i.e., $I_1=1$ and $I_2=1/2$), $\pi-$ (i.e., $I_1=-1$ and $I_2=1/2$), $\varphi-$ (i.e., $I_1=1$ and $I_2=-3/2$), and $\varphi_0-$ (i.e., $\varphi_0=0.4$) JJ (see the red, blue, purple, and black curves, respectively). For the first three cases refer to Eq.~\eqref{Chapter1:eqn:CPRphiJJ}, while for the last one to Eq.~\eqref{Chapter1:eqn:CPRphi0JJ}. (b) $(I_1,I_2)-$parameter space of the ground-state phase, $\tilde{\varphi}$, of the CPR in Eq.~\eqref{Chapter1:eqn:CPRphiJJ}.}
\label{Chapter1:fig:JosephsonEffect01}
\end{figure}

The first Josephson equation, $I_s (\varphi) = I_c \sin(\varphi)$, represents a particular case of the general Eq.~\eqref{Chapter1:eqn:GeneralCPR}.
A quite simple example of deviation from the simple sinusoidal CPR, in particular with the occurrence of the second harmonic sinusoidal contribution, is given by HTS \emph{d}-wave JJs and by junctions with ferromagnetic barriers~\cite{Barash1995,Tanaka1996,Buzdin2003,Stoutimore2018}. In this case:
\begin{equation}\label{Chapter1:eqn:CPRphiJJ}
I_s (\varphi) = I_1 \sin(\varphi) + I_2 \sin(2\varphi).
\end{equation}
From this CPR, the Josephson free energy can be written as
\begin{equation}
F_J (\varphi) = \frac{\Phi_0}{2\pi} \int_0^\varphi I_s (\varphi') d\varphi' = -\frac{\Phi_0}{2\pi} I_1\left[ \cos(\varphi) + \frac{I_2}{2I_1} \cos(2\varphi)\right].
\end{equation}
The ground state, $\tilde{\varphi}$, has to satisfy the condition
\begin{equation}
\left| \frac{dF_J (\varphi)}{d\varphi} \right|_{\varphi=\tilde{\varphi}} \propto I_s (\tilde{\varphi}) = 0.
\end{equation}
Since $ I_s (\varphi) = \sin(\varphi) [I_1 + 2I_2 \cos(\varphi)] $, the previous equation is satisfied if $ \sin(\tilde{\varphi}) = 0 $ or $ \cos(\tilde{\varphi}) = -\frac{I_1}{2I_2} $. This gives rise to the following cases:
\begin{enumerate}[itemindent=1cm,labelsep=1cm]
  \item If $ \left| \frac{I_1}{2I_2} \right| > 1 $ and $ I_2 > 0 $:
  \begin{itemize} [itemindent=0.5cm,labelsep=0.5cm]
  \item $ \tilde{\varphi} = 0 $ for $ I_1 > 0 $. If $ I_2 \in [0, 1] $ implies that $ I_1 \in [2I_2, 2] $;
  \item $ \tilde{\varphi} = \pi $ for $ I_1 < 0 $. If $ I_2 \in [0, 1] $ implies that $ I_1 \in [-2, -2I_2] $;
  \end{itemize}
  \item If $ \left| \frac{I_1}{2I_2} \right| < 1 $ and $ I_2 < 0 $:
  \begin{itemize} [itemindent=0.5cm,labelsep=0.5cm]
  \item $ \tilde{\varphi} = \pm \arccos\left(\frac{I_1}{2I_2}\right) $. Since $ I_2 < 0 $, we can assume $ I_2 \in [-1, 0] $, which implies $ |I_1| \in [0, 2|I_2|] $.
  \end{itemize}  
\end{enumerate}
These three cases are depicted in Fig.~\ref{Chapter1:fig:JosephsonEffect01}(a), see the red, blue, and purple curves, respectively. Moreover, Fig.~\ref{Chapter1:fig:JosephsonEffect01}(b) show the full $(I_1,I_2)$-parameter space of the ground-state phase, $\tilde{\varphi}$. The regions of $(I_1,I_2)$  values giving the $0-$, $\pi-$, and $\varphi-$junctions are evident. If $ \left| \frac{I_1}{2I_2} \right| < 1 $ but $ I_2 > 0 $, the potential profile shows, in addition to the ground state at $\tilde{\varphi}=0$ (if $I_1>0$) or $\tilde{\varphi}=\pi$ (if $I_1<0$), an extra metastable minimum at a higher energy at $\pi$ or $0$, respectively~\cite{Goldobin2007}.

The phase shift $ \tilde{\varphi} = \pi $ corresponds to an interesting case, i.e., a junction with $ I_c < 0 $, that is a negative supercurrent values for $ \varphi \in [0-\pi]$, the so-called \emph{$\pi-$junction}~\cite{Ryazanov2001,Buzdin2003,Goldobin2011}. Such a junction has an energy minimum at $ \varphi = \pi $, i.e., it provides a phase shift of $ \pi $ in the ground state, and may find a variety of applications in electronic circuits.
The crossover from $0-$ to $\pi-$state is observed in superconductor-ferromagnet-superconductor (SFS) junction.

The intrinsic phase shift $ \tilde{\varphi} = \pm \arccos\left(\frac{I_1}{2I_2}\right) $ corresponds to a two-fold degenerate state and $ I_s (\varphi) $ crosses the horizontal $\varphi$ axis at a position between $\varphi = 0$ and $\pi$. This case corresponds to the so-called \emph{$\varphi-$junctions}~\cite{Buzdin2003,Goldobin2015,Bakurskiy2018}. 

A different kind of unconventional junction, i.e., the so-called \emph{$\varphi_0$-junctions}~\cite{Bergeret2015,Szombati2016,Murani2017,Assouline2019,Mayer2020,Strambini2020,Guarcello2020,Idzuchi2021,Margineda2023,Maiellaro2024}, shows a non-trivial ($\varphi_0\neq0,\pi$) ground phase and is characterized by a significant even cosine component in the CPR, which transforms into
\begin{equation}\label{Chapter1:eqn:CPRphi0JJ}
I_s (\varphi) = I_c \sin(\varphi - \varphi_0),
\end{equation}
see the black curve in Fig.~\ref{Chapter1:fig:JosephsonEffect01}(a) for $\varphi_0=0.4$. Although, as we have seen, a non-trivial ground state can also emerge for the CPR in Eq.~\eqref{Chapter1:eqn:CPRphiJJ}, the CPR of a $\varphi_0-$junction gives a non-zero current at $\varphi=0$. In other words, this kind a junction can give a constant phase bias $ \varphi = -\varphi_0 $ in an open circuit configuration, or even a current $ I_s = I_c \sin(\varphi_0) $, i.e., the \emph{anomalous Josephson current}, if inserted into a closed superconducting loop.

Anomalous CPRs usually emerge in junctions under external magnetic or electric fields. It can also be obtained in the systems undisturbed by external interactions. For instance, SFS junctions with spin-orbit Rashba-type interaction demonstrate a significant shift of CPR, proportional to the product of the exchange field and the strength of spin-orbit coupling, but $\varphi_0-$junctions are also predicted in the systems with unconventional superconductors. Ferromagnetic anomalous JJs, resulting from joining superconductivity and ferromagnetic materials, have attracted considerable attention for their intriguing features, such as anomalous phase shifts and magnetization reversal induced by supercurrent flow, positioning them as promising candidates for potential applications in spintronics, quantum technologies, and information processing~\cite{Shukrinov2017,Guarcello2020b,Guarcello2021c,Shukrinov2022,Bobkova2022,Guarcello2023}. The appearance of this anomalous phase has been recently predicted even in a multiband scenario~\cite{Guarcello2022}.

Josephson systems with a non-trivial ground states, i.e., $\tilde{\varphi}$ or $\varphi_0$, allows the realization of phase shifter~\cite{Golod2021} and batteries~\cite{Pal2019,Strambini2020}. In fact, in a conventional JJ, the implementation of a phase battery is prevented by symmetry constraints, either time-reversal or inversion, which impose rigidity on the superconducting phase.
Conversely, the break of TRS ($t \rightarrow -t$) can lead to a phase shift 0 or $ \pi $, while the break of both TRS ($t \rightarrow -t$) and inversion symmetry ($r \rightarrow -r$) can lead to a phase shift $ \varphi_0 \in (0-\pi) $ and to an anomalous Josephson current, $ I_s (\varphi) = I_c \sin(\varphi - \varphi_0) $.

\subsection{Short JJ: the resistively and capacitively shunted junction (RCSJ) model}

\begin{figure}[t!!]
\centering{}\includegraphics[width=0.75\columnwidth]{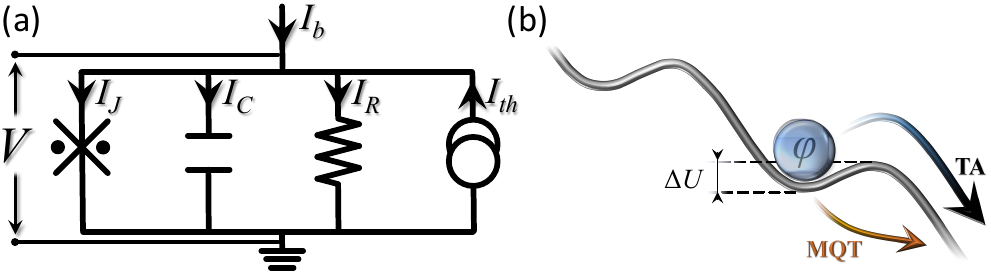}\caption{(a) Electrical model of a short JJ. The bias current, $I_b$, and the thermal noise current, $I_{th}$, are included in the diagram. (b) Sketch of the phase particle within a potential minimum of the tilted washboard potential $U$. The barrier height, $\Delta U$, is also shown.}
\label{Chapter1:fig:JosephsonEffect02}
\end{figure}

The CPR presented so far exclusively addresses the Cooper pair supercurrent. However, when $V >0$ and $T>0$, a quasiparticle current component also traverses the system. The complete time-dependent response of a JJ is appropriately captured by the equivalent circuit illustrated in Fig.~\ref{Chapter1:fig:JosephsonEffect02}(a), representing the so-called \emph{resistively and capacitively shunted junction} (RCSJ) model~\cite{Stewart1968,McCumber1968}. This circuit includes a capacitance $C = \epsilon_r \epsilon_0 A / t_{ox}$, where $A$ denotes the junction area, $t_{ox}$ and $\epsilon_r$ represent the thickness and relative permittivity of the oxide layer, and $\epsilon_0$ is the vacuum permittivity. The circuit also incorporates the resistance $R$ of the weak link. Owing to capacitive effects, tunnel junctions with typical high $R$ values exhibit hysteretic static IV characteristics. If the current surpasses the critical value, the JJ transitions to a state in which a measurable voltage drop appears, i.e., the so-called \emph{voltage state}. To mitigate the arising hysteresis, an additional shunt resistance $R_{sh}$ may be employed, resulting in an effective resistance equal to $RR_{sh}/(R+R_{sh})$.

The resistance $R$ exhibits nonlinear dependencies on both voltage and temperature~\cite{Likharev1986}, in particular, $R(V,T)=R_N$ if $V>V_g$ and $R(V,T)=R_{\text{sg}}(T)$ if $V\leq V_g$. , with $R_N$ and $R_{\text{sg}}$ being the normal-state and the subgap resistances~\cite{Tinkham2004}, respectively, and $V_g=2\Delta/e$ being the gap voltage (where $\Delta$ is the superconducting gap). However, being related to the dissipation in the system, its nonlinearity is often neglected in cases of moderate or weak damping. Alongside the resistor $R$, Fig.~\ref{Chapter1:fig:JosephsonEffect02}(a) shows a thermal noise current source, denoted as $I_{th}$, and the electric current, $I_b$, biasing the junction. Notably, the Josephson element can be alternatively viewed as a nonlinear inductance $L_J=L_c/\cos\varphi$, where $L_c=\Phi_0/(2\pi I_c)$.

Therefore, considering all current contributions and Josephson equations, we can easily derive the following Langevin equation using Kirchhoff's laws~\cite{Barone1982,BenJacob1984},
\begin{equation}
C\frac{\hbar}{2e}\frac{d^2\varphi}{dt^2}+\frac{1}{R}\frac{\hbar}{2e}\frac{d\varphi}{dt}+I_c\sin\varphi = I_b + I_{th}\left(t \right).
\end{equation}
By normalizing time with respect to the characteristic frequency $\omega_c=\frac{R}{L_c}=\frac{2e}{\hbar}\frac{I_c}{R}$, i.e., $\tilde{t}=\omega_c t$, and the current contributions on the rhs of this equation to the critical current $I_c$, previous equation becomes
\begin{equation}\label{Chapter1:eqJJ_norm_c}
\beta_c\frac{d^2\varphi}{d\tilde{t}^2}+\frac{d\varphi}{d\tilde{t}}+\sin\varphi = i_b + i_{th} (\tilde{t}),
\end{equation}
including the Stewart-McCumber damping parameter~\cite{Barone1982} $\beta_c=\frac{\omega_c^2}{\omega_p^2}$, with $\omega_p=\frac{1}{\sqrt{L_cC}}=\sqrt{\frac{2e}{\hbar}\frac{I_c}{C}}$ being the Josephson plasma frequency. 

Interestingly, Eq.~\eqref{Chapter1:eqJJ_norm_c} can even describe the damped motion of a particle in a tilted washboard-like potential~\cite{Barone1982}, see Fig.~\ref{Chapter1:fig:JosephsonEffect02}(b), expressed by 
\begin {equation}\label{Chapter1:eq:potential}
U(\varphi,i_b) = E_{J_0}[1 - \cos (\varphi) - i_b\; \varphi],
\end {equation}
where $E_{J_0}=\Phi_0/(2\pi)I_c$. The current $i_b$ biasing the system tilts $U(\varphi,i_b)$, so that for $i_b < 1$ the potential has metastable wells, with a barrier height equal to
\begin {equation}\label{Chapter1:eq:barrier}
\Delta\mathcal{U}(i_b)=\frac{\Delta U(i_b)}{E_{J_0}}=2 \left[ \sqrt{1-i_b^2} - i_b \cos^{-1}(i_b) \right].
\end {equation}
Instead, for $i_b \geq 1$ the potential profile has no maxima and minima.

The equation of the Josephson phase is also equivalent to the equation that governs the motion of a driven pendulum. Looking at a mechanical analog helps to better visualize the behavior of a JJ: for example, increasing the applied torque will increase the pendulum angle until it reaches $\pi/2$, at which point a minute additional applied twist, or noise, will cause the pendulum to flip over the top and then spin. Delving into this analogy, the phase difference $\varphi$ corresponds to the angle from vertical, the voltage $V$ to the angular velocity, the critical current $I_c$ to the restoring constant, the conductance $R^{-1}$ to the damping coefficient, the capacitance $C$ to the moment of inertia, and the bias current $I$ to the external torque.

\begin{figure}[t!!]
\centering{}\includegraphics[width=\columnwidth]{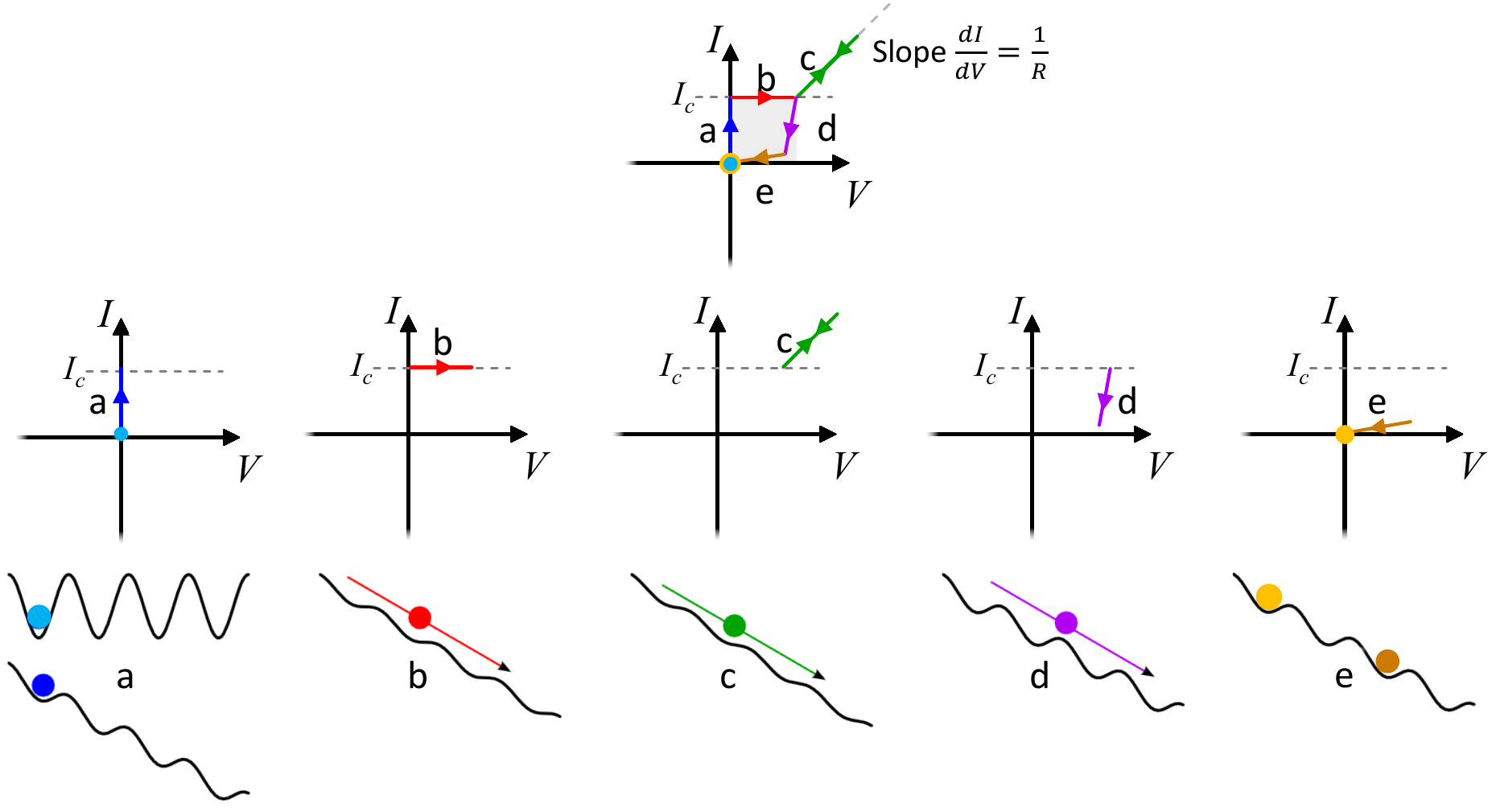}\caption{Scheme of the IV characteristic of a JJ. Top panel schematically sketches the whole path followed by the system when the bias current is first increased above $I_c$, and then reduced to zero. It can be divided in five distinct branches, represented in the bottom a-e panels. Each contains the corresponding characteristic IV branch and a sketch of the potential with the phase particle under stationary or moving (see the arrows) conditions.}
\label{Chapter1:fig:JosephsonEffect03}
\end{figure}

Viewing instead the problem in terms of a phase particle rolling on a rippled potential profile, in the presence of friction, helps to provide insight into the IV characteristic of a JJ. The latter is schematically illustrated in Fig.~\ref{Chapter1:fig:JosephsonEffect03}. We start with the system in the zero-voltage state, with the phase particle quite in a potential minimum. 
Then, increasing the bias current, the system remains in the zero-voltage state as long as $I<I_c$, see a). In this condition, the tilting imposed by the bias current to the washboard potential is not enough to vanish the potential barrier $\Delta U$, so that the phase particle remains confined in a potential well. When $I$ reaches $I_c$, $\Delta U =0$ and the particle starts to roll down along the potential profile, i.e., a non-zero voltage drop appears, see b). This holds even by increasing the current further, see c). Now, if the bias current is reduced, the system remains in the voltage state, at least as long as the current is greater than $I_c$. When $I_b<I_c$, the potential barriers appear again, but the following cases can occur:
\begin{itemize}
    \item Overdamped junction (where the friction term $\dot \varphi$ dominates): the combination of low kinetic energy and significant damping, makes the particle to rapidly stop, remaining trapped in a potential minimum, so that $\varphi$ ceases to evolve further. In this case, a non-hysteretic IV characteristic emerges;
    \item Underdamped junction (where the inertial term $\ddot \varphi$ dominates): the low friction and the high kinetic energy, makes the particle to continue moving down the tilted washboard potential, overcoming successive potential barriers. To stop the particle, the potential slope must be brought close to zero, requiring a very small $I_b$ to restore the zero voltage state. This phenomenon depends on the $\beta_c$ value. In this case, a hysteretic IV characteristic emerges.
\end{itemize}

The example case sketched in Fig.~\ref{Chapter1:fig:JosephsonEffect03} is strongly hysteretic, with a retrapping current much smaller than the critical current.
Finally, reducing further the bias current, the phase particle remains trapped again in a potential minimum, so the zero-voltage state is restored, and the system returns to the initial condition, see d) and e). 

The current value at which the system re-enters the zero-voltage state is called ``retrapping current''. Thus, in an overdamped system, switching and retrapping current coincide, $I_0=I_r$, whereas in an underdamped system they are different, and in particular $I_r/I_0\sim 4/(\pi\sqrt{\beta_c})$~\cite{Likharev1986}.

We have intentionally referred to ``switching'' and not critical current. In fact, if we assume to raise the bias current from zero and record the value at which the system switches to the voltage state, it turns out that this is below the expected critical current value, and that it is temperature-dependent. In fact, this mechanism is governed by thermal fluctuations, which are able to ``push'' the phase particle out of the metastable state even if the potential barrier is not completely zero, i.e., even if $I_b<I_c$. This point is the core of the next section.

\subsubsection{Temperature effects}

Thermal effects are accounted by the stochastic noise term, $I_{th}$, with the statistical properties
\begin{equation}\label{Chapter1:correl}
\langle I_{th}(t) \rangle = 0 \qquad\text{and} \qquad \langle I_{th}(t),I_{th}(t') \rangle = 2 \frac{k_BT}{R} \delta(t-t'),
\end{equation}
where $k_B$ is the Boltzmann constant, $\delta()$ is the Dirac delta function, and $T$ is the temperature. In normalized units, i.e., $\tilde{t}=\omega_c t$, previous equations become
\begin{equation}\label{Chapter1:correl_norm}
\langle i_{th}(\tilde{t}) \rangle = 0 \qquad\text{and} \qquad \langle i_{th}(\tilde{t}),i_{th}(\tilde{t}') \rangle = 2 \Gamma \delta(\tilde{t}-\tilde{t}'),
\end{equation}
where the adimensional noise intensity is
\begin{equation}\label{Chapter1:noiseampl}
\Gamma=\frac{k_B T}{E_{J_0}},
\end{equation}
i.e., the ratio between the thermal energy and the Josephson coupling energy $E_{J_0}$.

\begin{figure}[t!!]
\centering{}\includegraphics[width=0.49\columnwidth]{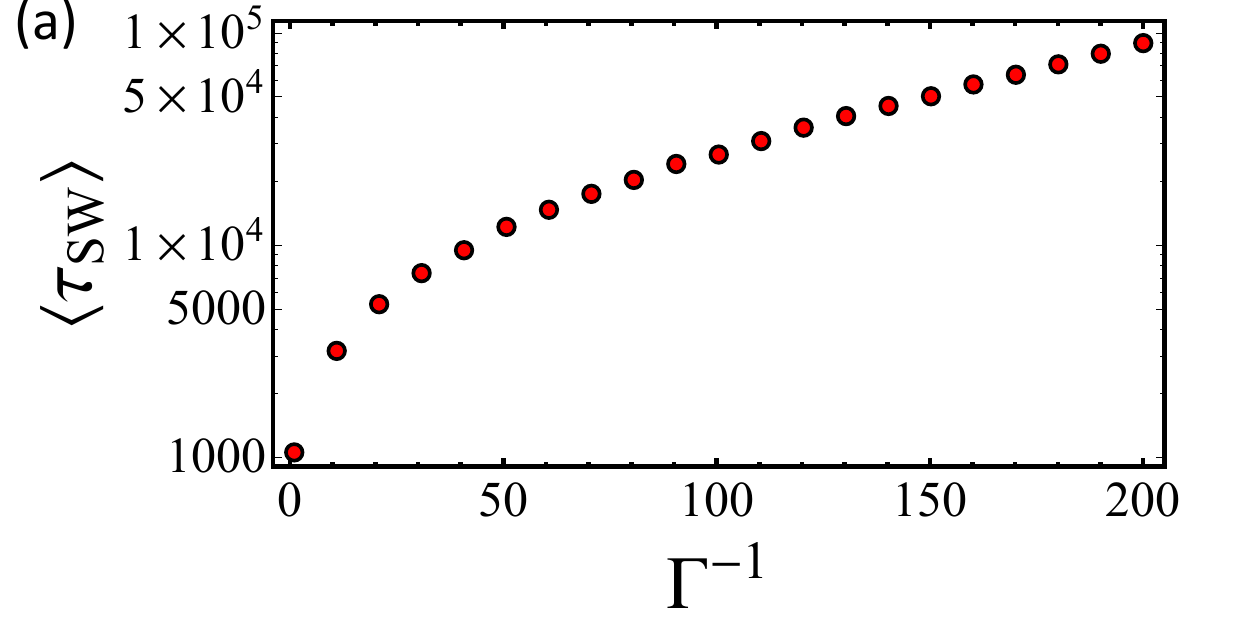}
\centering{}\includegraphics[width=0.49\columnwidth]{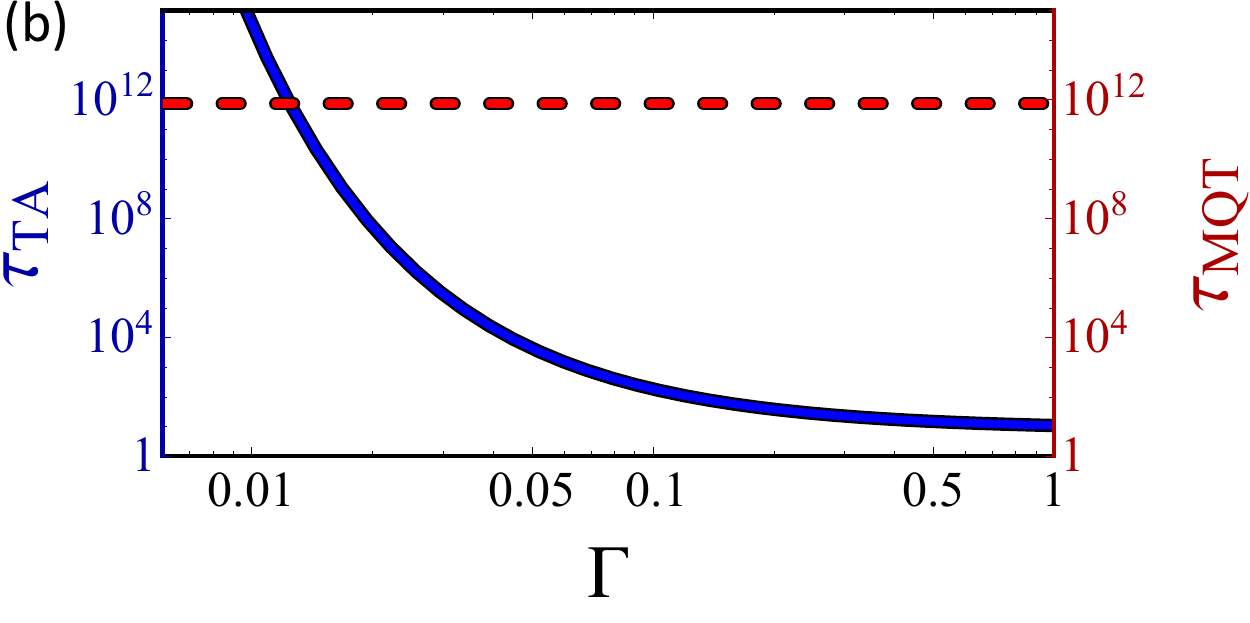}
\caption{(a) MST, $\tau_{\text{SW}}$ as a function of the inverse noise intensity, $\Gamma^{-1}$, obtained by numerical integration of Eq.~\eqref{Chapter1:eqJJ_norm_c}, setting $i_b=0.5$, $\beta_c=10^4$, and $N=10^4$. (b) TA ($\tau_{\text{TA}}$, left vertical axis, blue solid line) and MQT ($\tau_{\text{MQT}}$, right vertical axis, red dashed line) switching times, calculated as the inverse of the rates in Eqs.~\eqref{Chapter1:TAescape} and~\eqref{Chapter1:MQTescape}, as a function of the noise intensity $\Gamma$, at $i_b=0.7$ and $\beta_c=10^2$.}
\label{Chapter1:fig:JosephsonEffect04}
\end{figure}

The presence of the noise term makes escape from the metastable state an inherently stochastic process; therefore, when studying this type of dynamics, one typically reiterates the same ``experiment'' several times in order to construct distributions, from which to extract, e.g., mean values and variances, as, for instance, in the examples reported in Refs.~\cite{Guarcello2019,Guarcello2020c,Piedjou2021,Guarcello2021a,Guarcello2021b,Grimaudo2022,Grimaudo2023}. In particular, Fig.~\ref{Chapter1:fig:JosephsonEffect04}(a) shows the average switching time, obtained by averaging over $N=10^3$ independent repetitions, as a function of the inverse noise intensity, $\Gamma^{-1}$, setting $I_b=0.5$ and $\beta_c=10^4$.

As mentioned above, thermal fluctuations can drive the system out of a washboard potential minimum even in the case of a non-zero barrier height, i.e., for $I_b<I_c$. In fact, noise makes the phase solution metastable; thus the escape rate via thermal activation (TA), $\Gamma_{\textsc{ta}}$, is given by the Kramers approximation~\cite{Kramers1940}, which, for moderate damping, can be written as
\begin{equation}\label{Chapter1:TAescape}
 \Gamma_{\textsc{ta}}(\alpha,i_b,\Gamma) = \frac{1}{2\pi}\left( \sqrt{\frac{\alpha^2}{4}+\sqrt{1-i_b^2}}-\frac{\alpha}{2} \right) e^{-\frac{\Delta\, \mathcal{U}(i_b)}{ \Gamma}}.
\end {equation}
Interestingly, the prediction of an exponentially scaling average switching time for low noise is verified looking at the numerical results at small noise amplitudes in Fig.~\ref{Chapter1:fig:JosephsonEffect04}(a).

The phase particle can leave the metastable state also through a macroscopic quantum tunneling (MQT) mechanism, which rate is given by~\cite{Devoret1985}
\begin{equation}\label{Chapter1:MQTescape}
\Gamma_{\textsc{mqt}}(\alpha,i_b) =\frac{a_q}{2\pi} \sqrt[4]{1-i_b^2}e^{ \left [ -7.2\frac{\Delta U(i_b)}{\hbar \omega_p(i_b)}\left ( 1+0.87\alpha \right ) \right ]},\end{equation}
with $a_q=\sqrt{120\pi\left [ \frac{7.2 \Delta U(i_b)}{\hbar \omega_p(i_b)} \right ]}$, $\omega_p(i_b)=\omega_p\sqrt[4]{1-i_b^2}$, and $\alpha=1/\sqrt{\beta_c}$ is the damping parameter.
The MQT rate depends only on $\Delta U$, and not on $T$; this means that, as the temperature decreases, the TA rate can reduce so much that MQT processes dominate the escape dynamics. In this regard, one can define the so-called \emph{crossover temperature}~\cite{Grabert1984}, $T_{\text{cr}}$, as the temperature at which these two rates coincides: for $\alpha\ll1$ and $a_q\approx1$, it reduces to $T_{\text{cr}}\approx\frac{\hbar\omega_p}{7.2k_B}$. This is evident in Fig.~\ref{Chapter1:fig:JosephsonEffect04}(b), which shows the TA and MQT times, i.e., the inverse of the switching rates in Eqs.~\eqref{Chapter1:TAescape} and~\eqref{Chapter1:MQTescape}, $\tau_{\text{TA}}$ and $\tau_{\text{MQT}}$, respectively, as a function of the noise amplitude; in fact, the curves meet at $\Gamma\simeq0.012$, that corresponds to $T\simeq 0.012 \frac{\Phi_0}{2\pi k_B}I_c\simeq30\;\text{mK}$, assuming $I_c=0.1\; \mu \text{A}$.

\subsubsection{Magnetic field effects}

In this section, we focus on the effects produced by an applied magnetic field. So far, we looked at a so-called \emph{short} JJ (SJJ), specifically a junction where the magnetic field induced by the Josephson current is negligible in comparison to the externally applied magnetic field. Under these circumstances, the device dimensions are constrained to be smaller than the characteristic length-scale for such systems, defined as the \emph{Josephson penetration depth}~\cite{Barone1982}, which is derived in the following. 

\begin{figure}[t!!]
\centering{}\includegraphics[width=\columnwidth]{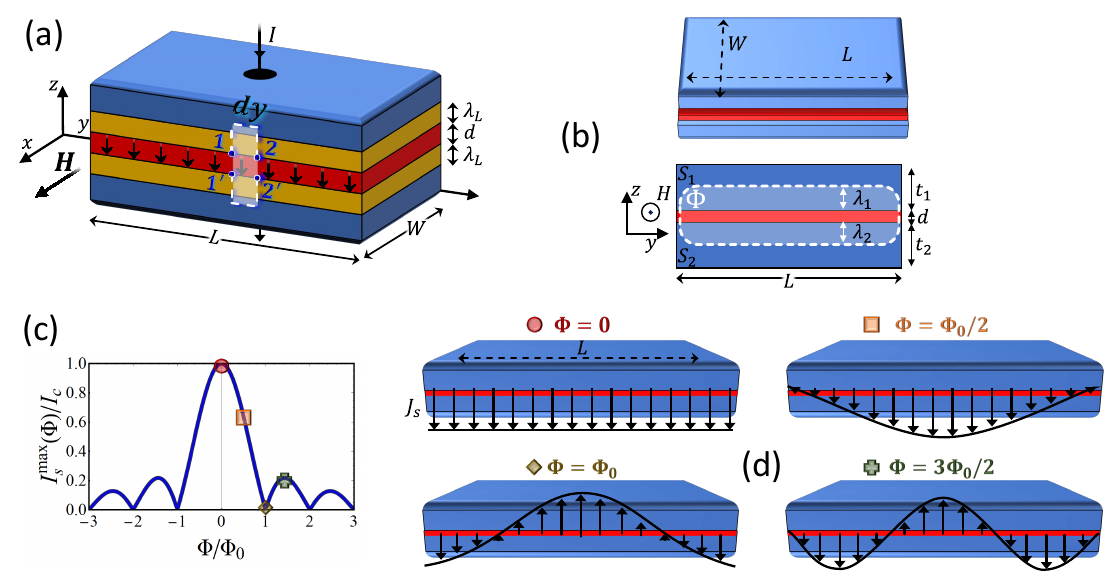}\caption{(a) Sketch of the SJJ, with indicated the path followed to perform the integrals in Eq.~\eqref{Chapter1:eqn:integralspath}. (b) Rectangular SJJ; the magnetic flux through the junction area is also reported. (c) Fraunhofer diffraction pattern for the critical current, with highlighted four situations of interest, corresponding to four different magnetic flux values. (d) Supercurrent density distributions along the junction length in the cases highlighted in (c).}
\label{Chapter1:fig:JosephsonEffect05}
\end{figure}

Let's assume a magnetic field applied in the junction plane and a rectangular barrier, like in Fig.~\ref{Chapter1:fig:JosephsonEffect05}(a). The magnetic flux density penetrates also into electrodes within a distance $t_H \approx d + 2\lambda_L$, i.e, the \emph{effective magnetic thickness} (see the yellow and red regions in the figure), assuming electrodes made by the same superconductors with a thickness larger than the London penetration depth $\lambda_L$ and $d$ being the insulating layer thickness. What is the relation between the magnetic field and the Josephson phase? We now that 
\begin{equation}
\textbf{B} = \nabla \times \textbf{A} \qquad\text{and}\qquad
\varphi = (\theta_2 - \theta_1) - \frac{2\pi}{\Phi_0} \int_{1}^{2} \textbf{A} \, dl.
\end{equation}

From Eq.~\eqref{Chapter1:eq:J_s}, one can obtain $\nabla \theta = \frac{2\pi}{\Phi_0} (\mu_0 \lambda_L^2 \textbf{J}_s + \textbf{A})$, where $\lambda_L = \sqrt{\frac{m^\star}{\mu_0 (q^\star)^2 n^\star}}$ and $\mu_0$ is the vacuum permeability. Then, we integrate $\nabla \theta$ along the white dashed lines in the two electrodes in Fig.~\ref{Chapter1:fig:JosephsonEffect05}(a), i.e.,
\begin{itemize}
\item Integration Path $2\;\to 1\;$: $\qquad\;\,\theta(1) - \theta(2)\; = \frac{2\pi}{\Phi_0} \mu_0 \lambda_L^2 \int_{2}^{1} \textbf{J}_s \, dl + \frac{2\pi}{\Phi_0} \int_{2}^{1} \textbf{A} \, dl$;
\item Integration Path $1'\to 2'$: $\qquad\theta(2') - \theta(1') = \frac{2\pi}{\Phi_0} \mu_0 \lambda_L^2 \int_{1'}^{2'} \textbf{J}_s \, dl + \frac{2\pi}{\Phi_0} \int_{1'}^{2'} \textbf{A} \, dl$.
\end{itemize}

Summing up these equations and including the integrals of $\textbf{A}$ across the barrier, one obtains:
\begin{equation}\label{Chapter1:eqn:integralspath}
\left[\theta(2') - \theta(2) - \frac{2\pi}{\Phi_0} \int_{2}^{2'}\!\!\! \textbf{A} dl \right] \!-\! \left[\theta(1') - \theta(1) - \frac{2\pi}{\Phi_0} \int_{1'}^{1}\!\!\! \textbf{A} dl \right] \!
= \! \frac{2\pi}{\Phi_0}\left[ \oint\!\!\! \textbf{A} dl + \mu_0 \lambda_L^2 \left(\int_{2}^{1} \!\!\!\textbf{J}_s dl + \int_{1'}^{2'}\!\!\! \textbf{J}_s dl\right)\right]\!.
\end{equation}
Since last term in the lhs vanishes, this equation reduces to:
\begin{equation}\label{Chapter1:eqn:phix1}
\frac{\partial\varphi}{\partial y} = \frac{2\pi}{\Phi_0} t_H B_x.
\end{equation}

According to this equation, an external magnetic field therefore affects the Josephson phase difference, producing measurable effects also in the critical current profile. Indeed, let's assume the geometry sketched in Fig.~\ref{Chapter1:fig:JosephsonEffect05}(b). 
If $\varphi(x,y)$ indicates the phase difference that is influenced by the external magnetic field, we have~\cite{Barone1982}:
\begin{equation}
\frac{\partial \varphi }{\partial y}=\frac{2\pi}{\Phi_0}t_H \mu_0 H_x(y) \qquad\text{and}\qquad \frac{\partial \varphi }{\partial x}=0.
\label{Chapter1:LocalMagneticField}
\end{equation}
These equations come from the condition $W\ll \lambda_J$, ensuring that $\varphi \left ( x,y \right )\equiv \varphi \left ( y \right )$. 

In a rectangular SJJ, since the external magnetic field is spatially homogeneous along the junction, i.e., $H_x(y)\equiv H_{ext}$, the phase exhibits a linear increase, which is given by 
\begin{equation}\label{Chapter1:eqn:phishort}
\varphi(y)=\left(2\pi\frac{\mu_0t_dH_{ext}}{\Phi_0} \right) y+\varphi_0=ky+\varphi_0.
\end{equation}
In contrast, in a \emph{long} junction, in which $L\gg \lambda_J$, both the penetrating external field and the self-field generated by the Josephson current contribute, resulting in a nonlinear variation of $\varphi(y)$ along the junction, see Eq.~\eqref{Chapter1:eqn:Ferrel-Prange}; this type of system will be faced in the next section. 

From Eq.~\eqref{Chapter1:eqn:phishort}, one obtains $J_s(x, y, t) = J_c(x, y) \sin(ky + \varphi_0)$, where the oscillation period of $J_s$ is $\Delta y = 2\pi/k = \Phi_0/(\mu_0 H_{ext} t_H)$. Thus, the magnetic flux through the junction within a single oscillation period corresponds to a single flux quantum: $\Phi\equiv\Delta y t_H \mu_0 H_{ext} = \Phi_0$.
The Josephson current can be calculated as:
\begin{equation}
I_s(H) \!=\! \int_{-\frac{L}{2}}^{\frac{L}{2}} \int_{-\frac{W}{2}}^{\frac{W}{2}} J_c(x, y) \sin(ky + \varphi_0) \, dx \, dy \!=\! \int_{-\frac{L}{2}}^{\frac{L}{2}} i_c(y) \sin(ky + \varphi_0) \, dy \!=\! \Im\left\{ e^{i\varphi_0} \int_{-\infty}^{\infty} i_c(y) e^{iky} \, dy \right\},
\end{equation}
where $i_c(y) = \int_{-W/2}^{W/2} J_c(x, y) \, dx$. In the last step we assumed $i_c(|y| > L/2) = 0$, so as to replace the integral limits with $\pm \infty$.
Moreover, if $i_c(|y| \leq L/2) = constant = i_c$, the maximum Josephson current becomes
\begin{equation}\label{Chapter1:eqn:Ismax1}
I_s^{\text{max}}(H) = \left| \int_{-\infty}^{\infty} i_c(y) e^{iky} \, dy \right|= \left|i_c \int_{-\frac{L}{2}}^{\frac{L}{2}} \cos(ky) \, dy\right| = \left|i_c \frac{1}{kL} \Big[ \sin(ky)\Big]_{-\frac{L}{2}}^{\frac{L}{2}}\right |= i_c L \left| \frac{\sin(kL/2)}{kL/2} \right|.
\end{equation}
Finally, since $kL/2 = \left(\frac{2\pi}{\Phi_0} \mu_0 H_x t_H\right) L/2 =\pi  \Phi/\Phi_0$ and $I_c = i_c L$, from Eq.~\eqref{Chapter1:eqn:Ismax1} we obtain a \emph{Fraunhofer diffraction pattern for the critical current}
\begin{equation}
\frac{I_s^{\text{max}}(\Phi)}{I_c} = \left| \frac{\sin\left(\pi \Phi/\Phi_0\right)}{\pi \Phi/\Phi_0} \right|,
\end{equation}
which is shown in Fig.~\ref{Chapter1:fig:JosephsonEffect05}(c). 
We stress that $I_c^{\text{max}}$ is not a current itself but only indicates the maximum value of the Josephson current. A different CPR and/or a different geometry can lead to a different magnetic-field dependence of the critical current, e.g., see Refs.~\cite{Weides2006,Kemmler2010,Suominen2017,Singh2022,Maiellaro2023,Maiellaro2024} and even a brief overview on some of the most significant deviations from the standard patterns given in~\cite{Tafuri2019}.

It is possible to make an intriguing parallel between the critical current distribution within a JJ and the renowned Fraunhofer diffraction pattern observed in optics. Let's start highlighting the $\Phi$-dependence of $\varphi(y)$, i.e.,
\begin{equation}\label{Chapter1:eqn:phishort2}
\varphi(y)=2\pi\frac{\mu_0t_dH_{ext}L}{\Phi_0}\frac{y}{L} +\varphi_0=2\pi\frac{\Phi}{\Phi_0}\frac{y}{L} +\varphi_0.
\end{equation}
Then, we focus on four situations of interest, corresponding to four different magnetic flux values, giving the critical currents highlighted in Fig.~\ref{Chapter1:fig:JosephsonEffect05}(c) with dedicated symbols. The resulting configurations of current density along the junction length are shown in Fig.~\ref{Chapter1:fig:JosephsonEffect05}(d). In particular, we look at the supercurrent density $i_c(y)$ and $J_s(x, y, t) = J_c(x, y) \sin(ky + \varphi_0)$:
\begin{itemize}
\item In the zero-field case, $\Phi=0$ so that $\varphi(y)=\varphi_0$ according to Eq.~\eqref{Chapter1:eqn:phishort2}, we have $i_c(y) = constant$. 
The maximum Josephson current (in the negative $z$ direction) is obtained for $\varphi_0=-\pi/2$, that is $J_s(x, y, t) = -J_c(x, y)$, see top-left panel of Fig.~\ref{Chapter1:fig:JosephsonEffect05}(d).
\item If $\Phi=\Phi_0/2$, $\varphi(y)=\pi y/L+\varphi_0$ and $\varphi(L/2)-\varphi(-L/2)=\pi$: since the supercurrent density oscillates sinusoidally, half of a full oscillation period lies within the junction. The choice of $\varphi_0$ says which ``half period'' is considered. In particular, in the top-right panel of Fig.~\ref{Chapter1:fig:JosephsonEffect05}(d) is shown the $\varphi_0=-\pi/2$ situation, which gives $\varphi(-L/2)=-\pi$, $\varphi(L/2)=0$ and the maximum Josephson current in the negative $z$ direction. 
\item If $\Phi=\Phi_0$, $\varphi(y)=2\pi y/L+\varphi_0$ and $\varphi(L/2)-\varphi(-L/2)=2\pi$, so that a full oscillation period lies within the junction, see bottom-left panel of Fig.~\ref{Chapter1:fig:JosephsonEffect05}(d). In this case, the total Josephson current is zero for each $\varphi_0$.
\item If $\Phi=3\Phi_0/2$, $\varphi(y)=3\pi y/L+\varphi_0$ and $\varphi(L/2)-\varphi(-L/2)=3\pi$, so that one and a half oscillation period lies within the junction, see bottom-right panel of Fig.~\ref{Chapter1:fig:JosephsonEffect05}(d). Current contributions from the full period are cancelled, so that the overall current is just given by the remaining half period. Naturally, the resulting current is less than that for $\Phi = \Phi_0/2$, where an entire semi-period is comprised into the system. This observation underscores the trend that the Josephson current typically diminishes as the applied magnetic field increases.
\end{itemize}

\subsection{Long JJ: the sine-Gordon model}

\begin{figure}[t!!]
\centering{}\includegraphics[width=\columnwidth]{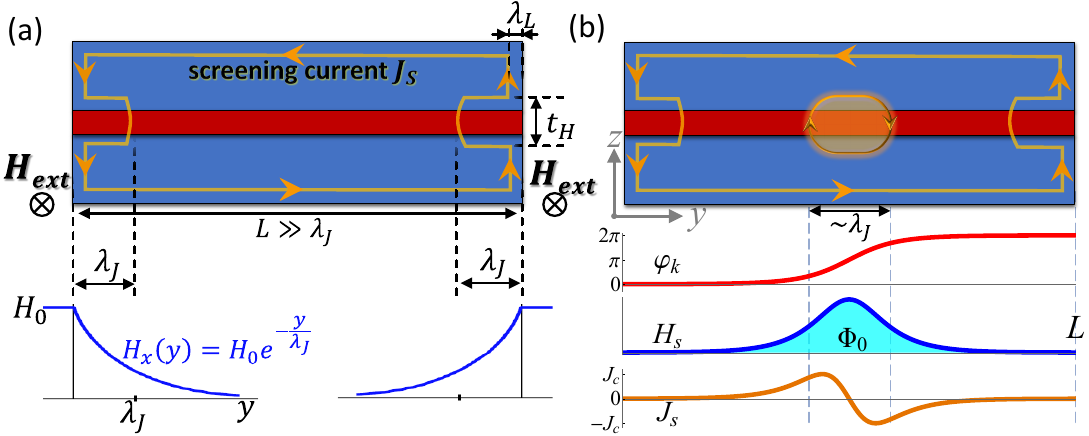}\caption{(a) Sketch of a LJJ in the case of a small applied field. Bottom panel shows the decaying magnetic field profile. (b) Sketch of a LJJ hosting a kink. Bottom panels shows a $2\pi$-phase profile and the corresponding magnetic field and supercurrent.}
\label{Chapter1:fig:JosephsonEffect06}
\end{figure}

In this section we consider a long JJ (LJJ) in which the barrier is in the $yz$-plane, the magnetic field is applied in $x$-direction resulting in phase variations along the $y$-direction, and the bias current is flowing in the negative $z$-direction, see Fig.~\ref{Chapter1:fig:JosephsonEffect06}. The magnetic flux density results both from the externally applied field and the Josephson current density and must satisfy Amp\'ere's law,
\begin{equation}\label{Chapter1:eqn:AmpereLaw}
\nabla \times \textbf{H} = \textbf{J} + \epsilon \epsilon_0 \frac{\partial \textbf{E}}{\partial t}
\qquad\text{that for our geometry becomes}\qquad
\frac{\partial H_x(y,t)}{\partial y} = -J_z(y,t) - \epsilon \epsilon_0 \frac{\partial E_x(y,t)}{\partial t}.
\end{equation}
If a spatial dimension, e.g., $L$, is larger than the Josephson penetration depth, we have to consider the self-field effect due the supercurrent. In the presence of an external magnetic field, we know that
\begin{equation}
\frac{\partial \varphi}{\partial y} = \frac{2\pi t_H \mu_0}{\Phi_0} H_x. \label{Chapter1:eq:flux_derivative}
\end{equation}
In the zero-voltage state, $\frac{\partial E}{\partial t} = 0$, Amp\'ere's law reduces to $\frac{\partial H_x}{\partial y} = -J_z(y)$. Since the current is flowing in the negative $z$-direction, $J_z(y) = -J_c \sin\varphi(y)$, from Eq.~\eqref{Chapter1:eq:flux_derivative}, we obtain
\begin{equation}\label{Chapter1:eqn:Ferrel-Prange}
\frac{\partial^2 \varphi}{\partial y^2} = \frac{2\pi t_H \mu_0}{\Phi_0} \frac{\partial H_x}{\partial y} = \frac{2\pi t_H \mu_0}{\Phi_0} J_c \sin\varphi(y) = \frac{1}{\lambda_J^2} \sin\varphi(y).
\end{equation}
This is often called the \emph{Ferrel-Prange equation}~\cite{Schmidt1997}, or more generally the \emph{stationary sine-Gordon equation} (SSGE). Here, we defined the Josephson penetration depth $\lambda_J = \sqrt{\frac{\Phi_0}{2\pi\mu_0 t_H J_c}}$: this is the characteristic length scale over which the magnetic field is screened (similar to $\lambda_L$ in a bulk superconductor). 

For small applied fields, Eq.~\eqref{Chapter1:eqn:Ferrel-Prange} becomes $\partial^2 \varphi/\partial y^2 = \varphi(y)/\lambda_J^2 $, giving a solution
\begin{equation}
\varphi(y) = \varphi(0) e^{-y/\lambda_J} \qquad\text{and}\qquad H_x(y) = -\frac{\Phi_0}{2\pi\lambda_J t_H \mu_0} \varphi(0) e^{-y/\lambda_J} = H_0 e^{-y/\lambda_J},
\end{equation}
where $\lambda_J$ clearly represents a decay length for the magnetic field, this justifying the expression ``penetration depth'', see Fig.~\ref{Chapter1:fig:JosephsonEffect06}(a).

In the case of $\frac{\partial E}{\partial t} \neq 0$, and considering that $E_x = -V/d$, $\frac{\partial \varphi}{\partial t} = \frac{2\pi V}{\Phi_0}$, and $J_x = -J_c \sin(\varphi)$, Eq.~\eqref{Chapter1:eqn:AmpereLaw} becomes
\begin{equation}
\frac{\partial^2 \varphi(y,t)}{\partial y^2} = \frac{2\pi}{\Phi_0} t_H \mu_0 \left[J_c \sin(\varphi(y,t)) + \epsilon \epsilon_0 \frac{\Phi_0}{2\pi d} \frac{\partial^2 \varphi(y,t)}{\partial t^2}\right].
\end{equation}
Using the definition of the Josephson penetration depth $\lambda_J$, one finally obtains the \emph{time-dependent Sine-Gordon equation} (SGE)~\cite{Barone1971,Parmentier1978,McLaughlin1978,Lomdahl1985,Ustinov1998,CuevasMaraver2014}
\begin{equation}
\frac{\partial^2 \varphi(y,t)}{\partial y^2} - \frac{1}{\bar{c}^2} \frac{\partial^2 \varphi(y,t)}{\partial t^2} - \frac{1}{\lambda_J^2} \sin(\varphi(y,t)) = 0,
\end{equation}
where $\bar{c} = \sqrt{\frac{d}{\epsilon \epsilon_0 t_H \mu_0}} = c\sqrt{\frac{1}{\epsilon(1+2\lambda_L/d)}}$ is the transverse electromagnetic velocity, called \emph{Swihart velocity}.
Recalling that $\omega_p = \sqrt{\frac{2\pi \Phi_0 I_c}{C_J}} = \sqrt{\frac{2\pi \Phi_0 I_c}{\epsilon \epsilon_0 A/d}} = \sqrt{\frac{2\pi \Phi_0 J_c d}{\epsilon \epsilon_0}}$, one obtains $\omega_p \lambda_J = \bar{c}$. Therefore, the SGE can be recast as
\begin{equation}
\lambda_J^2 \frac{\partial^2 \varphi(y,t)}{\partial y^2} - \frac{1}{\omega_p^2} \frac{\partial^2 \varphi(y,t)}{\partial t^2} - \sin(\varphi(y,t)) = 0.
\end{equation}

The SGE admits a kink solution, moving at a speed $v_y$ and centered in $y_0$, with expression:
\begin{equation}\label{Chapter1:eqn:kink}
\varphi_k(y,t) = 4 \arctan\left\{\exp\left[\pm\gamma_{\tilde{v}}\left(\frac{y-y_0}{\lambda_J} - \frac{v_y}{\bar{c}}t\right)\right]\right\}
\end{equation} 
where $\gamma_{\tilde{v}}=1\big/\sqrt{1-\tilde{v}^2}$, with $\tilde{v}=v_y/\bar{c}$. The sign $\pm1$ distinguishes a kink from an anti-kink. This solitonic solution corresponds to a $2\pi$-step in the phase profile, is stable, travels maintaining its shape (also after collision with other waves), gives rise to step structures in the IV characteristic, produces microwave radiation emission, and transports a clear magnetic flux equal to $\Phi_0$, see Fig.~\ref{Chapter1:fig:JosephsonEffect06}(b).

Equation~\eqref{Chapter1:eqn:kink} represents a fluxon moving with velocity $v_y$ along the junction. Under the action of the Lorentz force due to an externally applied current, the fluxon shifts along the junction, suffering Lorentz contraction on approaching the Swihart velocity $\bar{c}$.
The moving fluxon locally causes a temporal change of $\varphi$, which corresponds, according to the $2^{\text{nd}}$ Josephson relation, to a voltage pulse that becomes sharper with increasing velocity due to Lorentz contraction, in order to satisfy the condition $\int V dt = \Phi_0$. 

\begin{figure}[t!!]
\centering{}
\includegraphics[width=\columnwidth]{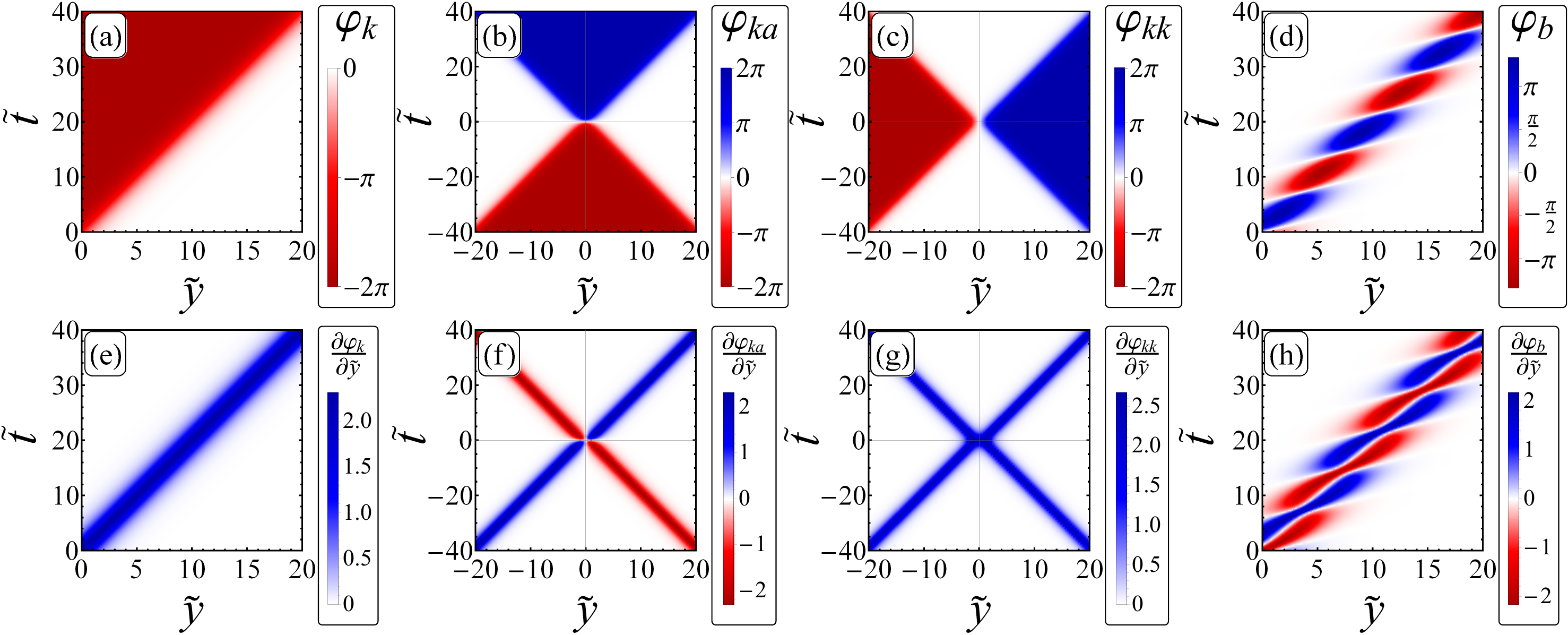}
\caption{Solutions of SGE (top panels) and corresponding space derivative (bottom panels): (a-e) kink $\varphi_k$, (b-f) kink-antikink collision $\varphi_{k-a}$, (c-g) kink-kink collision $\varphi_{k-k}$, and (d-h) breather $\varphi_b$. The other parameters are: $L=20$, $\tilde{v}=0.5$, and $\omega_b=0.5$.}
\label{Chapter1:fig:JosephsonEffect07}
\end{figure}

The SGE admits other kinds of solutions, e.g., (in normalized units and imposing $y_0=0$):
\begin{itemize}
\item kink-antikink collision
\begin{equation}
\varphi_{ka}(\tilde{y},\tilde{t}) = 4 \arctan\left\{\frac{1}{\tilde{v}}\frac{\sinh\left(\gamma_{\tilde{v}}\tilde{v}\tilde{t}\right)}{\cosh\left(\tilde{y}\gamma_{\tilde{v}}\right)}\right\};
\end{equation}
\item kink-kink collision
\begin{equation}
\varphi_{kk}(\tilde{y},\tilde{t}) = 4 \arctan\left\{\tilde{v}\frac{\sinh\left(\gamma_{\tilde{v}} \tilde{y}\right)}{\cosh\left(\tilde{v}\tilde{t}\gamma_{\tilde{v}}\right)}\right\};
\end{equation}
\item small amplitude oscillations (i.e., \emph{plasma waves}), in which case $\sin\left[\varphi(y,t)\right] \approx \varphi(y,t)$, so that
\begin{equation}
\lambda_J^2 \frac{\partial^2 \varphi(y,t)}{\partial y^2} - \frac{1}{\omega_p^2} \frac{\partial^2 \varphi(y,t)}{\partial t^2} = \varphi(y,t).
\end{equation}
This gives a solution
$\varphi(y,t) = \varphi_0 \exp\left[-i(ky-\omega t)\right]$, satisfying the dispersion relation $\frac{\omega}{\omega_p} = \sqrt{1+\left(\frac{k}{\lambda_J}\right)^2}$;
\item breather, with $\omega_b$ internal oscillation frequency,
\begin{equation}
\varphi_b(\tilde{y},\tilde{t}) = 4 \arctan\left\{\frac{\sqrt{1-\omega_b^2}}{\omega_b}\frac{\sin\left[\gamma_{\tilde{v}}\omega_b(\tilde{t}-\tilde{v} \tilde{y})\right]}{\cosh\left[\gamma_{\tilde{v}}\sqrt{1-\omega_b^2}(\tilde{y}-\tilde{v} \tilde{t})\right]}\right\}.
\end{equation}
\end{itemize}

Unlike kinks, the breather solution is unstable, oscillates periodically with time and decays exponentially in space, does not produce a potential difference, has no manifestations on the IV characteristic, does not produce a measurable magnetic flux through the JJ, must be generated efficiently and trapped in a confined area to allow measurements~\cite{Gulevich2012,DeSantis2022a,DeSantis2022b,DeSantis2023a,DeSantis2023b,DeSantis2023arXiv}.

Figure~\ref{Chapter1:fig:JosephsonEffect07} collects different solutions of SGE (top panels) and the corresponding space derivative (bottom panels): (a-e) a kink $\varphi_k$, (b-f) a kink-antikink collision $\varphi_{k-a}$, (c-g) a kink-kink collision $\varphi_{k-k}$, and (d-h) a breather $\varphi_b$. It is evident that the spatial derivative of the phase allows kinks and antikinks to be easily visualised, since each $2\pi$-step corresponds to a peak in the $ \partial\varphi/\partial x$ profile (in particular, positive for a kink and negative for an antikink). In this regard, Fig.~\ref{Chapter1:fig:JosephsonEffect07}(h) clearly demonstrates that a breather is composed of a bounded kink-antikink pair oscillating at a proper internal frequency.

\begin{figure}[t!!]
\centering{}\includegraphics[width=0.5\columnwidth]{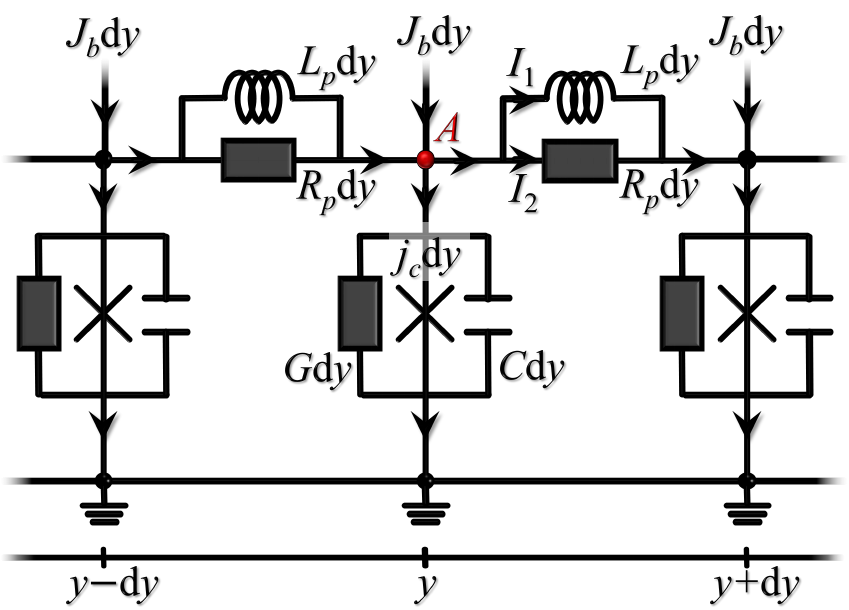}\caption{Equivalent circuit elements constituting the LJJ analog~\cite{Scott1976,Lomdahl1982}.}
\label{Chapter1:fig:JosephsonEffect08}
\end{figure}

In analogy to short junctions, an equivalent circuit can also be designed for LJJs~\cite{Scott1976,Lomdahl1982}, see Fig.~\ref{Chapter1:fig:JosephsonEffect08}. Focusing on the node $A$, applying Kirchhoff current law one obtains
\begin{equation}
\frac{\partial (I_1+I_2)}{\partial y} = J_b - C \frac{\partial V}{\partial t} - \frac{V}{R} - J_c \sin(\varphi).
\end{equation}
Here, $L_p$ is the inductance per unit length representing the magnetic energy stored within one $\lambda_L$ of the superconducting film, and $R_p$ is a resistance per unit length representing the scattering of quasiparticles in the surface layers of the two superconductors. Using the $2^{\text{nd}}$ Josephson relation and the relation $dV = -R_p I_2 dy = -L_p \frac{\partial I_1}{\partial t} dy$ one obtains
\begin{equation}
\frac{\partial I_1}{\partial t} = -\frac{1}{L_p} \frac{dV}{dy} = -\frac{1}{L_p} \frac{\Phi_0}{2\pi} \frac{\partial^2 \varphi}{\partial y \partial t},
\end{equation}
where
\begin{equation}
I_1 = -\frac{1}{L_p} \frac{\Phi_0}{2\pi} \frac{\partial \varphi}{\partial y} \qquad\text{and}\qquad
I_2 = -\frac{1}{R_p} \frac{dV}{dy} = -\frac{1}{R_p} \frac{\Phi_0}{2\pi} \frac{\partial^2 \varphi}{\partial y \partial t},
\end{equation}
so that
\begin{equation}
-\frac{1}{L_p} \frac{\Phi_0}{2\pi} \frac{\partial^2 \varphi}{\partial y^2} - \frac{1}{R_p} \frac{\Phi_0}{2\pi} \frac{\partial^3 \varphi}{\partial y^2 \partial t} = J_b - C \frac{\Phi_0}{2\pi} \frac{\partial^2 \varphi}{\partial t^2} - \frac{1}{R} \frac{\partial \varphi}{\partial t} - J_c \sin(\varphi).
\end{equation}
Then, by normalizing $\tilde{y} = y/\lambda_J$ and $\tilde{t} = \omega_p t$ (note that another choice for normalising time has been made previously; in fact, it is usual to normalise times to the inverse of the plasma or the characteristic frequency), one obtains the \emph{perturbed SGE} in normalized units
\begin{equation}\label{Chapter1:eqn:perturbed SGE}
\frac{\partial^2 \varphi}{\partial \tilde{y}^2} - \frac{\partial^2 \varphi}{\partial \tilde{t}^2} = \sin(\varphi) + \alpha \frac{\partial \varphi}{\partial \tilde{t}} - \beta \frac{\partial^3 \varphi}{\partial \tilde{y}^2 \partial \tilde{t}} - \gamma,
\end{equation}
where $\alpha = 1/({R C \omega_p})$ (tunneling of quasiparticles), $\beta = \omega_p L_p/R_p$ (surface currents damping), and $\gamma = J_b/J_c$ (normalized bias current density).

Equation~\eqref{Chapter1:eqn:perturbed SGE} is completed by boundary conditions taking into account the external magnetic field, $H_{ext}(t)$:
\begin{equation}
\frac{\partial \varphi(0,t)}{\partial y} = \frac{2\pi t_H \lambda_J}{\Phi_0} H_{ext}(t) \qquad\text{and}\qquad
\frac{\partial \varphi(L,t)}{\partial y} = \frac{2\pi t_H \lambda_J}{\Phi_0} H_{ext}(t).
\end{equation}

At a small $H_{\text{ext}}$ value, we can imagine a sort of ``tail'' of a fluxon, which is ideally centred outside the junction, penetrating the junction itself. Then, increasing the magnetic field, the center $y_0$ of this fluxon moves closer to the edge of the junction, according to~\cite{Goldobin2001}
\begin{equation}
\frac{2}{\cosh(y_0/\lambda_J)} = \frac{2\pi}{\Phi_0} H_{\text{ext}} \lambda_J = \frac{H_{\text{ext}}}{H_0},
\end{equation}
where $H_0 = \Phi_0 / (2\pi t_{\text{H}} \lambda_J)$. At $H_{\text{ext}} = H_{c_1} = 2H_0$, we have that $y_0$ vanishes, i.e., a fluxon is set at each edge the junction. At $H_{\text{ext}} > H_{c1}$, fluxons penetrate the LJJ from the edges and fill it with some density which depends on the value of $H_{\text{ext}}$.

Looking at the behavior of the screening current, for small applied fields the junction can screen the applied external field by a circulating screening current, which flows in the opposite direction at both junction edges and adds to the external applied bias current. Increasing the applied field, the screening current enlarges until it reaches the critical value at one edge. Then vortices start to penetrate the junction resulting in an oscillating $J_{\text{S}}(z)$ dependence. In particular, the state with no vortex in the junction is called the \emph{Meissner state}, while for magnetic fields larger than the critical field $H_{c_1}$, one or more vortices enter the junction.

The magnetic field dependence of critical current density in LJJs leads to diffraction patterns reminiscent of the ``Fraunhofer-like'' phenomena. In the context of SJJ limits, the diffraction lobes are typically well-separated; however, in the case of LJJs we observe a overlapping lobe structure in the $J_c(H)$ pattern. The interpretation of these patterns involves the insertion of kinks into the junction. Each individual lobe corresponds to a state characterized by a specific and unchanging number of kinks. As the magnetic field strength increases, configurations with a higher kinks number become more energetically favorable: in this case, the system undergoes a transition from a metastable state to a more stable one featuring an increased number of kinks. Within the range of magnetic field strengths where diffraction lobes overlap, different solutions with different $N$ may concurrently coexist, but the system stays in the most energetically stable. 

Looking at thermal effects, even in the case of LJJ these can be included in Eq.~\eqref{Chapter1:eqn:perturbed SGE} by a stochastic noise term, with the statistical properties
\begin{equation}\label{Chapter1:correlLJJ}
\langle J_{th}(y,t) \rangle = 0 \qquad\text{and} \qquad \langle J_{th}(y,t),J_{th}(y,t') \rangle = 2 \frac{k_BT}{R} \delta(t-t') \delta(y-y'),
\end{equation}
that in normalized units, i.e., $\tilde{y} = y/\lambda_J$ and $\tilde{t} = \omega_p t$, become
\begin{equation}\label{Chapter1:correl_norm_2}
\langle j_{th}(\tilde{y},\tilde{t}) \rangle = 0 \qquad\text{and} \qquad \langle j_{th}(\tilde{y},\tilde{t}),j_{th}(\tilde{y},\tilde{t}') \rangle = 2 \Gamma \delta(\tilde{t}-\tilde{t}')\delta(\tilde{y}-\tilde{y}'),
\end{equation}
where the noise intensity now is given by $\Gamma=\frac{k_B T}{R}\frac{\omega_p}{\lambda_J J_c^2}$. In the literature, there are many examples of long Josephson systems in which the effects of noise are effectively taken into account to better explain the device's response, e.g., see Refs.~\cite{Fedorov2007,Fedorov2008,Fedorov2009,Augello2009,Pankratov2012,Valenti2014,Guarcello2015,Guarcello2016,Pankratov2017}.

Finally, we conclude mentioning that a mechanical analog can be given even for a LJJ~\cite{Barone1971,Barone1982}, that is a chain of coupled pendula with torque. In particular, the $\varphi_{xx}$ term is represented by the restoring torque of the spring, the $\varphi_{tt}$ term by the moments of inertia f the pendula, and the $\sin(\varphi)$ by the gravitational torque. 

\subsection{The superconducting quantum interference device (SQUID)}

\begin{figure}[t!!]
\centering{}\includegraphics[width=0.8\columnwidth]{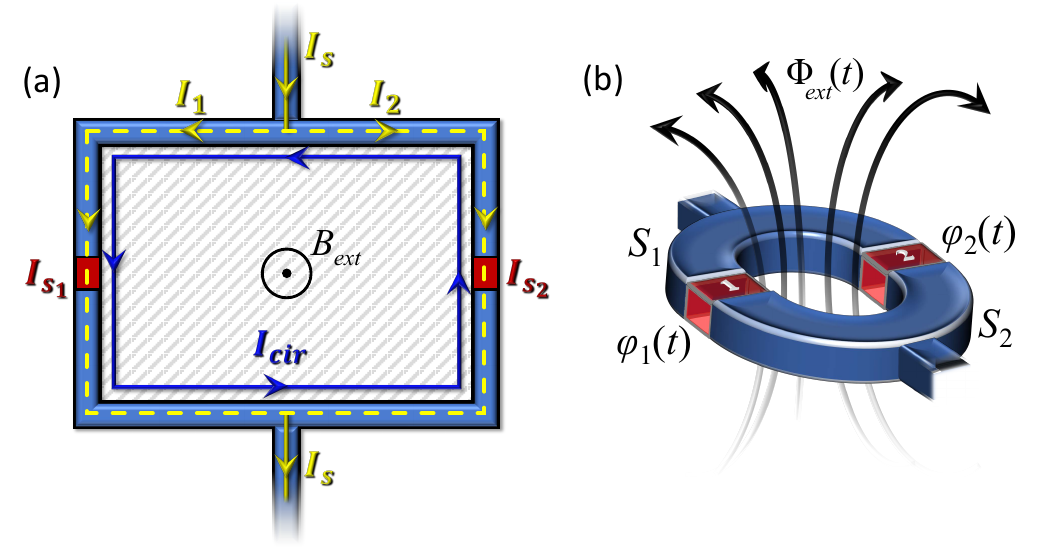}\caption{Sketches of a two-junction dc-SQUID, including the circulating and the flowing current through the SQUID loop.}
\label{Chapter1:fig:JosephsonEffect09}
\end{figure}

Two junctions interrupting a superconducting loop constitute a \emph{direct-current superconducting quantum interference device} (SQUID), i.e., a \emph{dc-SQUID}, see Fig.~\ref{Chapter1:fig:JosephsonEffect09}. The name suggests the use of an external current biasing the SQUID. For a description of the main properties and usual applications of SQUIDs, see Refs.~\cite{Clarke2004,Granata2016}.

Considering an integration path within the superconducting loop, see the yellow dashed line in Fig.~\ref{Chapter1:fig:JosephsonEffect09}(a), the total flux, $\Phi$, through the ring, taking into account also the Josephson phase drop across each JJ, can be written as:
\begin{equation}
2\pi \frac{\Phi}{\Phi_0} = \oint_{C} \Delta\theta \, dl = 2\pi n + \varphi_2 - \varphi_1.
\end{equation}
The working principles of a SQUID combines two physical phenomena: the flux quantization in the superconducting loop and the Josephson effect. SQUIDs are the most sensitive detectors for magnetic flux $\Phi$, being essentially a Flux-to-Voltage converter giving a flux-dependent output voltage with a period of one flux quantum, $\Phi_0$.

The total current flowing through the SQUID is
\begin{equation}
I_s = I_{s1} + I_{s2} = I_{c_1} \sin(\varphi_1) + I_{c_2} \sin(\varphi_2).
\end{equation}
If, for simplicity, one assumes $I_{c_1} = I_{c_2} = I_c$, the total current can be written as
\begin{equation}
I_s = 2I_c \cos\left(\frac{\varphi_1 - \varphi_2}{2}\right) \sin\left(\frac{\varphi_1 + \varphi_2}{2}\right)= 2I_c \cos\left(\pi \frac{\Phi}{\Phi_0}\right) \sin(\phi),
\end{equation}
by defining $\phi = (\varphi_1 + \varphi_2)/2$.
If the inductance of the superconducting loop is negligible, the total flux coincides with the external flux, $\Phi = \Phi_{\text{ext}}$: in this case, the maximum supercurrent can be written as
\begin{equation}
I_s^{\text{max}} = 2I_c \left|\cos\left(\pi \frac{\Phi_{\text{ext}}}{\Phi_0}\right)\right|.
\end{equation}
However, if the inductance $L$ of the superconducting loop is non-negligible, we obtain
\begin{equation}
\frac{\Phi}{\Phi_0} = \frac{\Phi_{\text{ext}}}{\Phi_0} + \beta_L \cos(\phi) \sin\left(\pi \frac{\Phi}{\Phi_0}\right),
\end{equation}
where $\beta_L = 2LI_c/\Phi_0$ is the \emph{screening parameter}, which is one of the most important parameters, since the SQUID characteristic strongly depends on the $\beta_L$ value~\cite{Clarke2004}.
In what follows, we look at $I_s^{\text{max}}(\Phi)$ when $I_{c_1}$ and $I_{c_2}$ are the same or different and $\beta_L$ is zero or not: 
\begin{itemize}
\item If $I_{c_1} \neq I_{c_2}$ and $\beta_L = 0$, defining $\alpha=I_{c_2}/I_{c_1}$, the maximum supercurrent can be written as
\begin{equation}
I_s^{\text{max}}(\Phi_{\text{ext}}) = I_{c_1} \sqrt{(1-\alpha)^2 + 4\alpha \cos^2\left(\pi \frac{\Phi_{\text{ext}}}{\Phi_0}\right)}.
\end{equation}
\item If $I_{c_1} = I_{c_2}$ and $\beta_L = 0$
\begin{equation}
I_s^{\text{max}}(\Phi_{\text{ext}}) = 2I_c \left|\cos\left(\pi \frac{\Phi_{\text{ext}}}{\Phi_0}\right)\right|.
\end{equation}
\item If $I_{c_1} = I_{c_2} = I_c$ and $\beta_L \neq 0$, one has to solve numerically the equations
\begin{equation}\label{Chapter1:eqn:SQUIDequations1}
I_s = 2I_c \cos\left(\pi \frac{\Phi}{\Phi_0}\right) \sin(\phi) \qquad\text{and}\qquad
\frac{\Phi}{\Phi_0} = \frac{\Phi_{\text{ext}}}{\Phi_0} + \beta_L \cos(\phi) \sin\left(\pi \frac{\Phi}{\Phi_0}\right).
\end{equation}
\end{itemize}

\begin{figure}[t!!]
\centering{}\includegraphics[width=0.49\columnwidth]{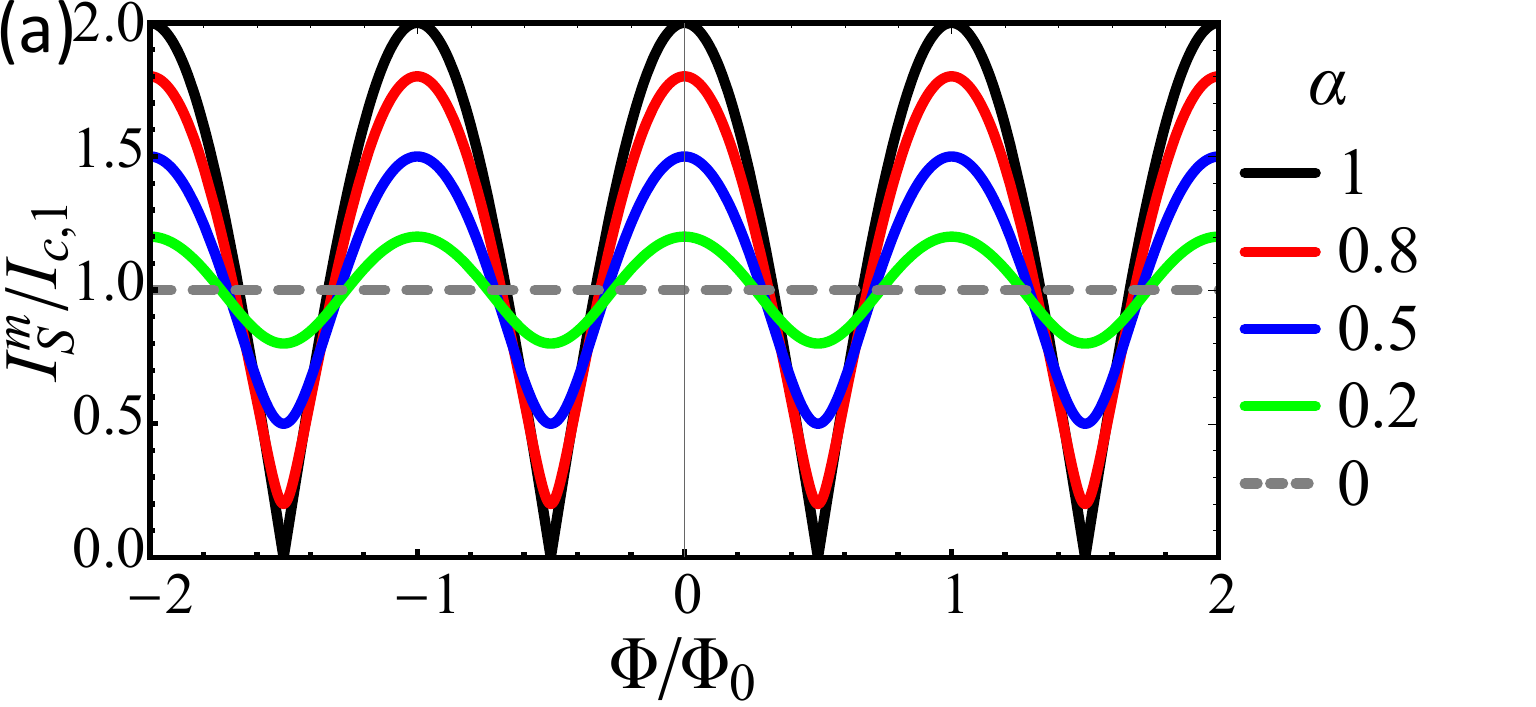}
\centering{}\includegraphics[width=0.49\columnwidth]{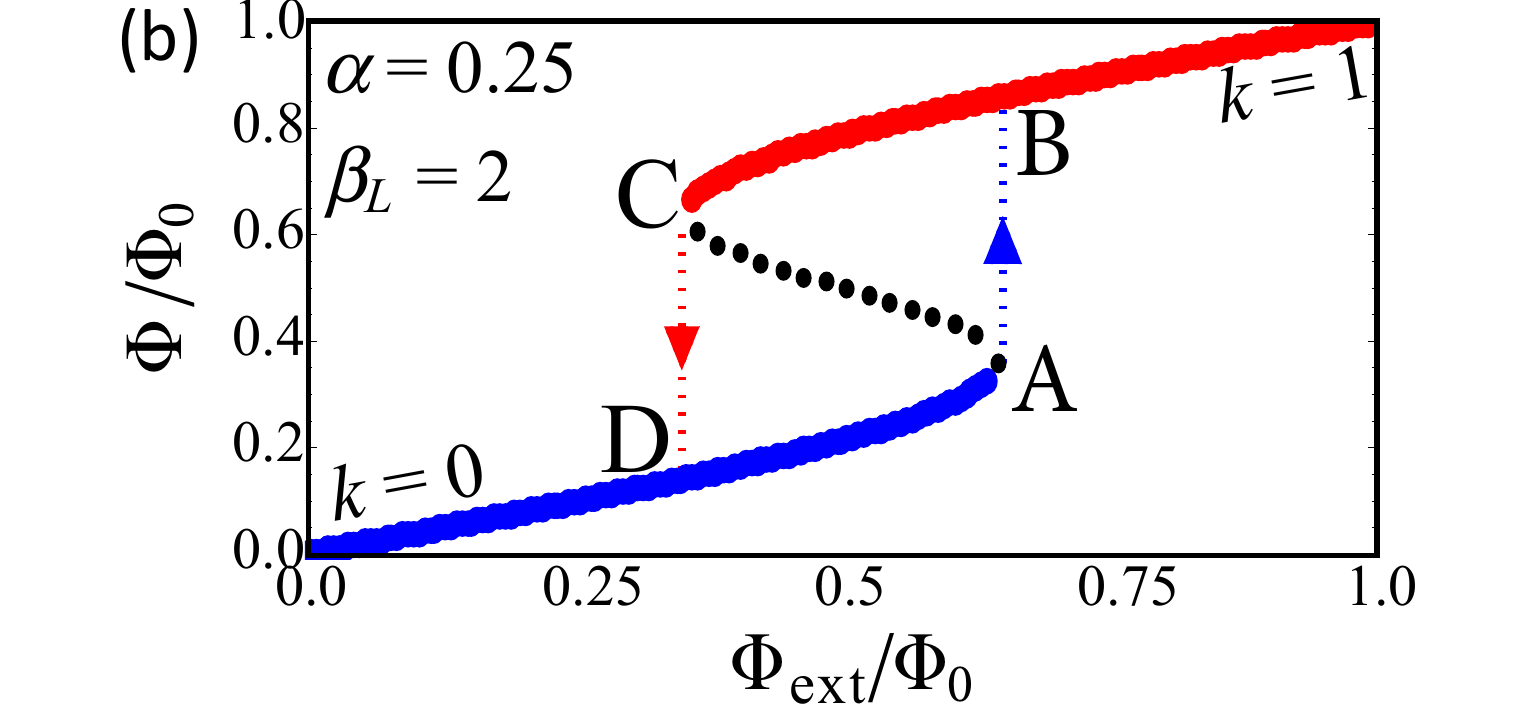}
\caption{(a) Critical current interference pattern of a two-junction SQUID as a function of the critical current ratio $\alpha=I_{c,2}/I_{c,1}$. (b) Normalized total magnetic flux through the SQUID loop, $\Phi/\Phi_0$, as a function of the normalized external flux, $\Phi_{\text{ext}}/\Phi_0$. Arrows indicate the transition between quantum states with different numbers $k$ of flux quanta. The dotted part of the curve represents unstable states. Labels A, B, C, and D serve to mark a hysteresis loop followed by the total magnetic flux when sweeping $\Phi_{\text{ext}}$ back and forth.}
\label{Chapter1:fig:JosephsonEffect10}
\end{figure}

In the case of identical currents (and very low inductance) the normalized critical current modulates between 2 and 0, see black curve in Fig.~\ref{Chapter1:fig:JosephsonEffect10}(a). Increasing the asymmetry between the JJs, that is reducing $\alpha$, causes the modulation of the oscillation of the critical current pattern to be progressively reduced, until it is cancelled if the critical current of a junction is zero; in other words, two junctions are required to see an interference pattern, see Fig.~\ref{Chapter1:fig:JosephsonEffect10}(a). A reduction of the modulation depth occurs even for nonzero values of $\beta_L$. For $\beta_L = 1$ the critical current modulates by 50$\%$, and for $\beta_L \gg 1$, the modulation $\Delta I_c/I_{c,\text{max}}$ decreases as $1/\beta_L$~\cite{Clarke2004}. In principle, one could use the strong $I_c(\Phi_{\text{ext}})$ modulation to operate the SQUID as a sensitive flux detector. However, this process takes some time, so it is simpler to operate the SQUID in the voltage state, in which one sends a current $I_b$ and measures the voltage drop $V$~\cite{Granata2016}.

The SQUID can operate in the so-called \emph{hysteretic mode}~\cite{Clarke2004,Guarcello2017,Guarcello2018}. The circulating current $I_{\text{cir}}$, see the blue line in Fig.~\ref{Chapter1:fig:JosephsonEffect09}(a), tends to compensate the applied flux $\Phi_{\text{ext}}$; in particular, for a sufficiently high $\beta_L$ value, the total flux can become multivalued, and abrupt flux transitions occur when changing $\Phi_{\text{ext}}$. This is schematically shown in Fig.~\ref{Chapter1:fig:JosephsonEffect10}(b).
With increasing $\Phi_{\text{ext}}$, the total flux follows a branch with a positive slope; we observe that only the parts with a positive slope are associated to stable states of the system, i.e., the dotted branch in Fig.~\ref{Chapter1:fig:JosephsonEffect10}(b) is unstable. If we increase $\Phi_{\text{ext}}$ further, at a certain moment the supercurrent flowing in the SQUID reaches the critical current of the weak link. Then, the junction momentarily switches to the voltage state, and the SQUID undergoes a $k\rightarrow k+1$ transition (the magnetic flux through the SQUID changes by one flux quantum). 
Now, if the applied flux is decreased, at a certain value a $k+1\rightarrow k$ transition occurs. In other words, by sweeping $\Phi_{\text{ext}}$ forth and back, the system undergoes transitions from one state to another and back, so that a hysteresis loop (like ABCDA in the figure) is traced on the phase diagram, see Fig.~\ref{Chapter1:fig:JosephsonEffect10}(b).

According to the RCSJ model, the phase evolution of a current-biased dc-SQUID formed by two identical JJs (i.e., the symmetric case) is determined by the following normalized equations of motion, with $k=1,2$ and $\varphi_i$ being the Josephson phase of the $k$-th junction:
\begin{equation}
\beta_c \frac{\partial^2 \varphi_k}{\partial \tilde{t}^2} + \frac{\partial \varphi_k}{\partial \tilde{t}} = -\frac{\partial \mathcal{U}_{\textsc{SQUID}} }{\partial \varphi_k},
\end{equation}
where
\begin{equation}
\mathcal{U}_{\textsc{SQUID}} = \frac{U_{\textsc{SQUID}}}{E_{J_0}} = \frac{2}{\pi \beta_L} \underbrace{\left(\frac{\varphi_2 - \varphi_1}{2} - \pi \varphi_a\right)^2}_{\text{Magnetic Energy}} + \underbrace{\Big(-\cos(\varphi_1) - \cos(\varphi_2)\Big)}_{\text{Josephson Energy}} - \frac{i(\varphi_2 + \varphi_1)}{2}.
\end{equation}
Here, $i$ is the normalized bias current, $j$ is the normalized circulating current, and we have defined
\begin{equation}
\beta_L = \frac{E_{J_0}}{E_M}, \qquad\text{where}\qquad E_{J_0} = \frac{\Phi_0 I_c}{2\pi}, \quad\text{and}\quad E_M = \frac{1}{2\pi} \left(\frac{\Phi_0^2}{2L}\right).
\end{equation}

We can generalize the RCSJ equations to the case of different JJs (asymmetric case), where $C_k$, $R_k$, and $I_{c_k}$ are the capacitance, the normal resistance, and the critical current of the $k$-th junction~\cite{Clarke2004}. Moreover, $\Phi = \Phi_{\text{ext}} + L I_{\text{cir}}$, where $L = L_1 + L_2$ (with $L_1$ and $L_2$ being the inductances of the two SQUID arms) is the inductance of the superconducting ring. By defining $ \bar{I}_c = \frac{1}{2}(I_{c_1} + I_{c_2})$, $R = \frac{2R_1 R_2}{R_1 + R_2}$, and $ \bar{C} = \frac{1}{2}(C_1 + C_2)$, in the case of
\begin{align}
I_{c_1} \neq I_{c_2}, \quad R_1 \neq R_2, \quad C_1 \neq C_2, \quad\text{and}\quad L_1 \neq L_2,
\end{align}
one can define asymmetry parameters $\alpha_I$, $\alpha_R$, $\alpha_C$, and $\alpha_L$, so that
\begin{align}
I_{c_1} &=  \bar{I}_c (1 - \alpha_I), \qquad R_1 = \frac{R}{1 - \alpha_R}, \qquad C_1 = \bar{C}(1 - \alpha_C), \qquad L_1 = \frac{L(1 - \alpha_L)}{2} \\
I_{c_2} &=  \bar{I}_c (1 + \alpha_I), \qquad R_2 = \frac{R}{1 + \alpha_R}, \qquad C_2 = \bar{C}(1 + \alpha_C), \qquad L_2 = \frac{L(1 + \alpha_L)}{2}.
\end{align}
By including also two independent Johnson-Nyquist noise currents, $I_{n_1}$ and $I_{n_2}$, with the usual white noise features,
\begin{equation}
\langle I_{n_i}(t) \rangle = 0\qquad\text{and}\qquad
\langle I_{n_i}(t) I_{n_i}(t') \rangle = 2 \frac{k_B T}{R_i} \delta(t - t'),
\end{equation}
the RCSJ model in normalized units can be therefore written as:
\begin{equation}
\begin{aligned}
\frac{i}{2} + j &= (1 - \alpha_I) \sin(\varphi_1) + (1 - \alpha_R) \frac{\partial \varphi_1}{\partial \tilde{t}} + \beta_c (1 - \alpha_C) \frac{\partial^2 \varphi_1}{\partial \tilde{t}^2} + i_{n_1} \\
\frac{i}{2} - j &= (1 + \alpha_I) \sin(\varphi_2) + (1 + \alpha_R) \frac{\partial \varphi_2}{\partial \tilde{t}} + \beta_c (1 + \alpha_C) \frac{\partial^2 \varphi_2}{\partial \tilde{t}^2} + i_{n_2}.
\end{aligned}
\end{equation}
The total flux is given by 
\begin{equation}
\Phi_T = \Phi_{\text{ext}} + L_1 I_1 - L_2 I_2 = \Phi_{\text{ext}} + L_J - \alpha_L L I/2
\end{equation}
and
\begin{equation}
\varphi_2 - \varphi_1 = 2\pi \Phi_{\text{ext}}/\Phi_0 + \pi \beta_L (j + \alpha_L i/2).
\end{equation}

Regarding the thermal fluctuations~\cite{Clarke2004}, they can become relevant when the thermal energy, $k_B T$, approaches:
\begin{itemize}
\item the Josephson coupling energy, $E_{J_0}$, i.e.,
\begin{equation}
\Gamma = \frac{k_B T}{E_{J_0}} =\frac{2\pi/\Phi_0 k_B T}{I_c} =\frac{I_{\text{th}}}{I_c} \rightarrow 1.
\end{equation}
\item the characteristic magnetic energy, $E_M = E_{J_0}/\beta_L = \Phi_0^2/(2L)/(2\pi)$, i.e., 
\begin{equation}
\Gamma \beta_L = \frac{k_B T}{E_M} =\frac{L}{\Phi_0^2/(4\pi k_B T)} =\frac{L}{L_{\text{th}}} \rightarrow 1.
\end{equation}
\end{itemize}

On the contrary, a small thermal fluctuations regime is established when:
\begin{itemize}
\item $\Gamma \ll 1$, that means $I_c \gg I_{\text{th}} = \frac{2\pi}{\Phi_0} k_B T \propto T$. To give a realistic number, $I_{\text{th}} \sim 0.18 \;\mu\text{A}$ at $T = 4.2\;\text{K}$.
\item $\Gamma \beta_L \ll 1$, that means $L \ll L_{\text{th}} = \Phi_0^2/(4\pi k_B T) \propto 1/T$. To give a realistic number, $L_{\text{th}} \sim 5.9\;\text{nH}$ at $T = 4.2\;\text{K}$.
\end{itemize}

Let's look now at the so-called \emph{radio-frequency SQUID}, i.e., an \emph{rf-SQUID}, which consists of a superconducting loop interrupted by one or more JJs, with no externally applied bias current, and inductively coupled to an LC resonant circuit (i.e., a tank circuit)~\cite{Clarke2004}. In the case of a single-junction rf-SQUID (for a description of a double-junction rf-SQUID, see Ref.~\cite{Bo2004}), the fluxoid quantization condition gives $2\pi\Phi/\Phi_0+2\pi n=-\varphi$. The circulating supercurrent and the total magnetic flux through the SQUID loop satisfy
\begin{equation}
I_s=-I_c\sin\left(2\pi\frac{\Phi}{\Phi_0}\right)
\qquad\text{and}\qquad
\frac{\Phi}{\Phi_0}
=\frac{\Phi_a}{\Phi_0}
-\frac{\beta_{L,\text{rf}}}{2\pi}
\sin\left(2\pi\frac{\Phi}{\Phi_0}\right),
\end{equation}
where $\Phi_a$ is the total applied flux, including the contribution induced by the tank circuit, and $\beta_{L,\text{rf}}=2\pi LI_c/\Phi_0$ is the screening parameter, with $L$ denoting the SQUID loop inductance.

The SQUID loop is inductively coupled to the inductor $L_T$ of a parallel LC tank circuit, characterized by the quality factor $Q=R_T/(\omega_{\text{rf}}L_T)$, where $R_T$ is the equivalent parallel resistance and $\omega_{\text{rf}}=1/\sqrt{L_TC_T}$ is the uncoupled resonance frequency. When driven at resonance by an rf-current $I_{\text{rf}}\sin(\omega_{\text{rf}}t)$, the unloaded tank circuit supports a circulating current of amplitude $I_T\simeq QI_{\text{rf}}$. In the following, we discuss the response of the tank circuit for both $\beta_{L,\text{rf}}<1$ and $\beta_{L,\text{rf}}>1$.

For $\beta_{L,\text{rf}}<1$, the SQUID response is non-hysteretic, and the device behaves as a nonlinear, flux-dependent inductor. The flux linked to the tank inductor is
\begin{equation}
\Phi_T=L_TI_T+MI_s,
\end{equation}
where $M=\alpha\sqrt{L_TL}$ is the mutual inductance, with $\alpha$ denoting the dimensionless coupling coefficient. The total applied flux threading the SQUID loop is $\Phi_a=\Phi_{\text{ext}}+MI_T$, where $\Phi_{\text{ext}}$ is the externally applied magnetic flux in the absence of the tank current. Therefore, the total flux through the SQUID loop is
\begin{equation}
\Phi=\Phi_{\text{ext}}+MI_T+LI_s.
\end{equation}

In the small-signal, quasistatic limit, the SQUID response can be linearized around a stationary operating point. Differentiating the flux relation with respect to $I_T$, at fixed $\Phi_{\text{ext}}$, gives
\begin{equation}
\frac{d\Phi}{dI_T}
=M+L\frac{dI_s}{dI_T}.
\end{equation}
Using the Josephson current-phase relation, we obtain
\begin{equation}
\frac{dI_s}{d\Phi}
=-\frac{2\pi I_c}{\Phi_0}
\cos\left(2\pi\frac{\Phi}{\Phi_0}\right)
=-\frac{\beta_{L,\text{rf}}}{L}
\cos\left(2\pi\frac{\Phi}{\Phi_0}\right).
\end{equation}
Combining these expressions yields
\begin{equation}
\frac{dI_s}{dI_T}
=-\frac{M}{L}
\frac{\beta_{L,\text{rf}}\cos(2\pi\Phi/\Phi_0)}
{1+\beta_{L,\text{rf}}\cos(2\pi\Phi/\Phi_0)}.
\end{equation}

The effective differential inductance of the tank circuit is therefore
\begin{equation}
L_{T,\text{eff}}
=\frac{d\Phi_T}{dI_T}
=L_T+M\frac{dI_s}{dI_T}
=L_T\left[
1-\alpha^2
\frac{\beta_{L,\text{rf}}\cos(2\pi\Phi_{\text{eq}}/\Phi_0)}
{1+\beta_{L,\text{rf}}\cos(2\pi\Phi_{\text{eq}}/\Phi_0)}
\right].
\end{equation}
Here, $\Phi_{\text{eq}}$ denotes the equilibrium flux threading the SQUID loop, obtained from Eq.~(85) in the absence of the rf excitation ($I_T=0$). Consequently, the effective differential inductance varies periodically with the externally applied flux. In the small-signal, quasistatic approximation, neglecting additional dissipative loading, the corresponding tank-circuit resonance frequency is $\omega_{\text{rf,eff}}=1/\sqrt{L_{T,\text{eff}}C_T}$. When the tank circuit is driven close to resonance, this flux-dependent frequency shift produces a modulation of its rf-current and rf-voltage response.

For $\beta_{L,\text{rf}}>1$, the situation is different, since the relation $\Phi(\Phi_{\text{ext}})$ can become multivalued, leading to hysteresis and flux jumps. The tank-circuit response then depends on the nonlinear dynamics of the coupled system, and the small-signal expression for $L_{T,\text{eff}}$ derived above is not generally sufficient to describe its operation. Nevertheless, the tank voltage can still exhibit a periodic dependence on the applied magnetic flux.

\section{Josephson devices}

In this part of the chapter we will make an overview of the design and technologies associated to circuits involving JJs, and similia. We start with showing how JJs are made and how material properties and technical requirements determine the technologic solution implemented.

\subsection{JJ fabrication technology}

As already stated, JJs are made by properly joining two pieces of superconducting materials. This can be made in a variety of different ways. However, the needs to realize useful circuits with the junctions implies that a planar technology, based on thin films of suitable materials, has to be implemented. This is because most of other circuit elements, such as resistors, inductors, and capacitors, have been implemented in planar technology for the realization of integrated circuits.
The choice of the superconductor to be used depends mainly on the application foreseen and on the available technology. Historically, superconducting elements with low melting point (the so-called soft materials, such as Aluminium, Lead, Indium and Thin) were used, due to the relatively simple thin-film deposition techniques and to the relatively easy fabrication of an insulating layer separating two films. These materials were easy to fabricate, but have a relatively low critical temperature. A breakthrough in the technology was achieved when full $Nb$ JJs were realized using a in situ trilayer technique, opening the possibility to realize reliable and durable superconducting circuits operating at the liquid Helium temperature of 4.2K. This has been the real starting for the development of superconducting electronics. The discovery of compound superconducting materials having higher critical temperatures, such as $NbN$, $Nb_{3}Sn$, $BKBO$, $MgB_2$, and the cuprates, above all the $YBCO$, triggered the interest in developing new technologies for the realization of reliable and easier to use, from the point of view of the involved cryogeny, superconducting circuits. However, with the notable exception of $NbN$ and $YBCO$ for different applications, all these new materials showed specific difficulties related to the realization of JJs that impeded the realization of useful circuits.
In Table 1 is reported a list of different superconducting materials used to realize superconducting circuits, organized with increasing critical temperature and divided in zones corresponding to very low, low, medium, and high critical temperature superconductors (VLTS, LTS, MTS, HTS). Beside the critical temperature other material parameters, relevant for applications, such as the London penetration length and the coherence length, are reported.

\begin{table}
    \centering
    \begin{tabular}{|c|c|c|c|c|} \hline 
         Class&  Material&  $\lambda_L$ [nm]&  $\xi$ [nm]& Tc [K]\\ \hline 
         VLTS&  $Al$&  16&  1500& 1.18\\ \hline 
         VLTS&  $In$&  25&  400& 3.3\\ \hline 
         VLTS&  $Sn$&  28&  300& 3.7\\ \hline 
         LTS&  $Pb$&  28&  110& 7.2\\ \hline 
         LTS&  $Nb$&  32&  39& 9.2\\ \hline 
         MTS&  $NbN$&  50 (200)&  6& 17\\ \hline 
         MTS&  $Nb_{3}Sn$&  50&  6& 18\\ \hline 
         MTS&  $MgB_2$&  140&  3-5& 39\\ \hline 
         HTS&  $YBCO$&  140&  1.5& 92\\ \hline
    \end{tabular}
    \caption{Superconducting materials relevant for electronic applications organized in increasing critical temperature.}
    \label{Chapter1:tab:materiali}
\end{table}

In order to have a clean thin film of material, there are different deposition techniques that can be used. Those employed for superconductors are mainly: thermal evaporation, e-beam evaporation, and sputtering. A detailed description of these techniques is beyond the scope of this chapter. The interested reader is referred to the book from Milton Ohring~\cite{ohring2001materials}. The type of deposition technique, and the deposition parameters, strongly affects the properties of the deposited thin films, their crystalline structure, the eventual gas inclusions, the resulting stress, which determine the lattice constant and ultimately their transport and superconducting properties. Niobium, the workhorse of superconducting electronics, is particularly sensitive to the deposition conditions. 
After the deposition of a thin film, in order to build the wanted circuits, it is necessary to pattern it into a defined geometrical form. The related technologies, generally knows as photolithography, are very similar to those employed in semiconductor industry. They are based on sequences of photo- (or electron-) sensitive layer deposition (called photo-resist), its exposure (to UV or e-beam), its development and excess film material etching.

\begin{figure}[b!!]
\centering{}\includegraphics[width=0.75\columnwidth]{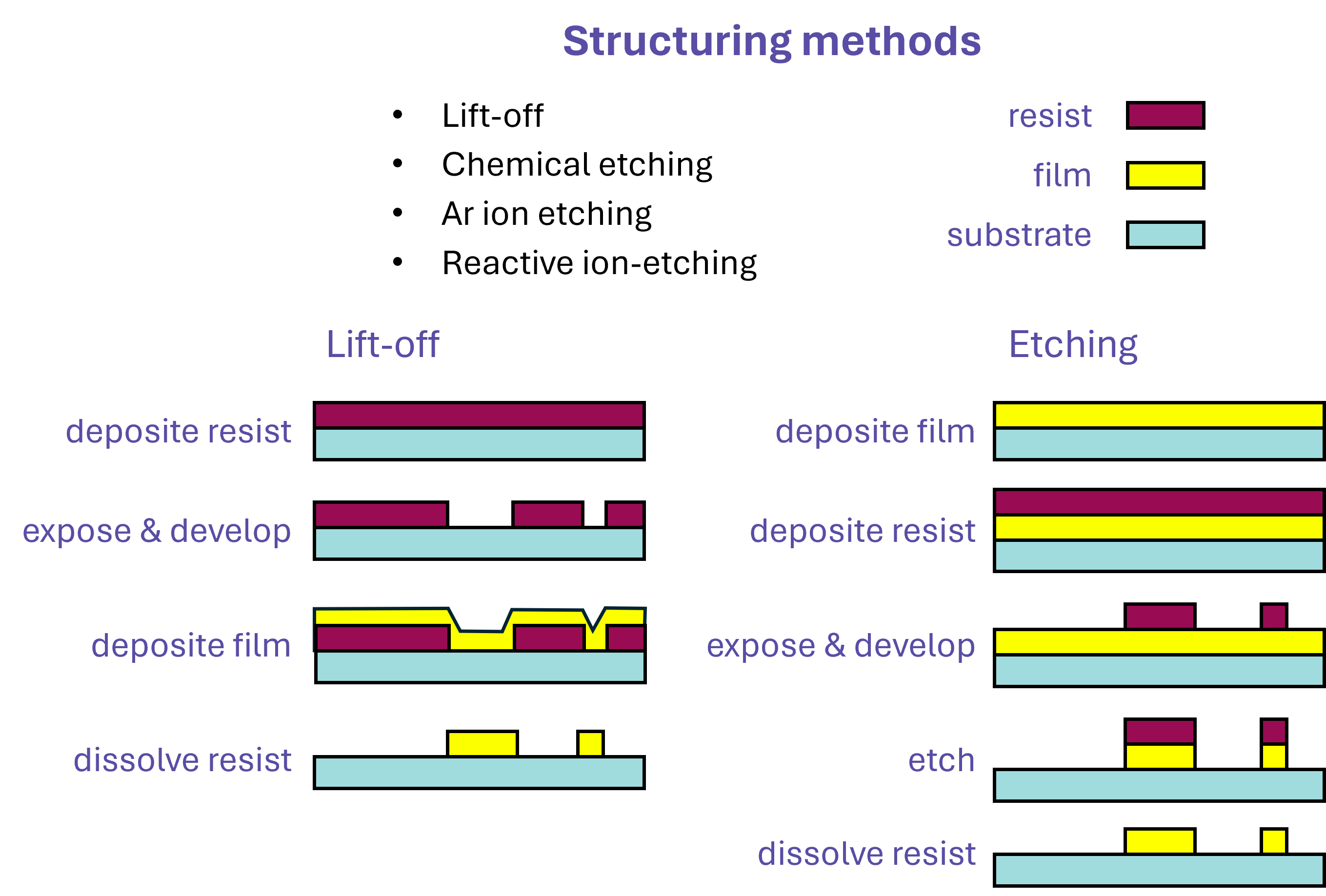}\caption{Typical thin film structuring methods}
\label{Chapter1:fig:etching}
\end{figure}

In Fig.~\ref{Chapter1:fig:etching} two different commonly used structuring methods are shown: lift-off and direct etching. The etching of the film can be made in different ways: lift-off, wet chemical etching, physical ion etching and reactive ion etching.
The lift-off technique essentially consists in predefining, with the resist, of the negative image of the wanted geometry, followed by the deposition of the thin film. By subsequently dissolving the resist in an appropriate solvent, also the thin film on its top is removed, leaving the wanted geometrical pattern on the substrate.
The direct etching techniques are somehow opposite. First the thin film is deposited on all the substrate area, it is subsequently covered with a thin layer of resist and then the desired geometry is defined using photolithographic techniques. The remaining exposed part of the film can be removed in different ways: using and appropriate chemical solution that dissolves the material through a chemical reaction (chemical etching), using a beam of energetic particles, typically Ar ions, that impinging on the film with enough energy, physically remove it (physical etching), or using a reactive plasma, typically $CF_4$ or $SF_6$, that chemically reacts with the material producing volatile reaction products, thus eroding the exposed material.
Each technique has its pros and cons, in terms of complexity, effects on the material to be patterned, effects on the quality of the patterning, and in particular on the achievable limit resolution.
For example, the lift-off technique has the advantage of being gentle with respect to the thin film manipulation and allows the use of shadow evaporation technique~\cite{dolan1977offset}, which is a key technology for the realization of superconducting qubits. 

The realization of a JJ-based device using a planar thin film technology requires a number of fabrication steps in order to realize the wanted coupling between two superconductors in a well defined position and with well defined geometrical shapes. During the several decades of development of JJ technology, different processes have been developed and optimized. Practically all the materials listed in Table~\ref{Chapter1:tab:materiali}, and some more, have been used to realize JJs for specific applications. However, nowadays most of the electronic applications of JJs are bases on Nb as electrode material and on the Nb trilayer technology for the junction fabrication~\cite{gurvitch1983high}. A notable exception is the realization of superconducting qubits which, with their specific needs in terms of operating temperature, stray capacitance and low dissipation, are currently mostly realized using the $Al$ as main material. These latter two will be discussed here in some detail.

\begin{figure}[b]
    \centering
    \includegraphics[width=0.75\linewidth]{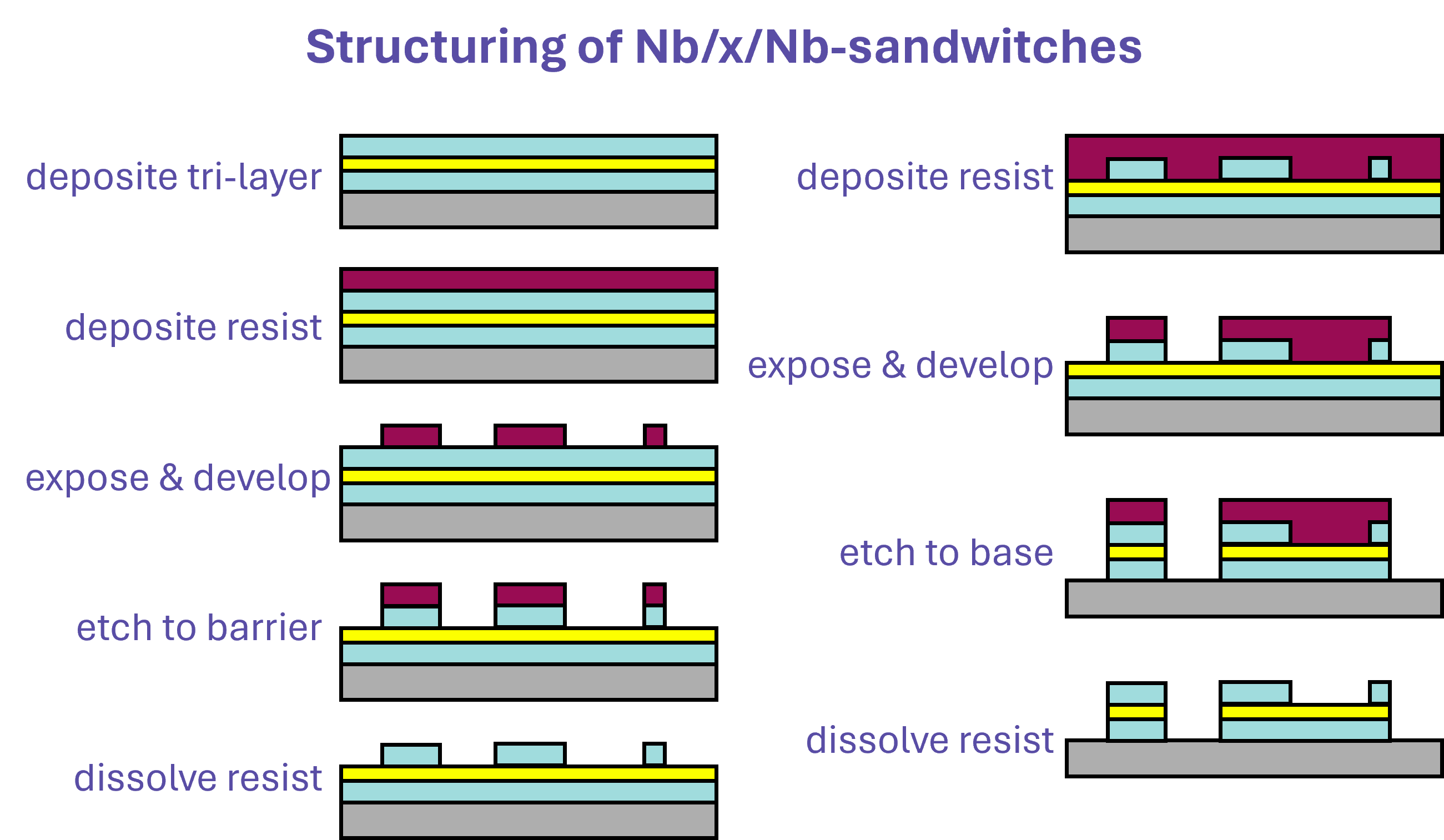}
    \caption{Some steps from the sequence for the fabrication of a $Nb$-based JJ using the trilayer technique. The sequence reads from top to down and from left to right.}
    \label{Chapter1:fig:trilayer}
\end{figure}

In Fig.~\ref{Chapter1:fig:trilayer} it is shown a typical sequence used for the fabrication of $Nb$-based JJs using the trilayer technique. The grey area represents the substrate, typically a $Si$ wafer covered with a thin ($200-300\;\text{nm}$) insulating layer of $SiO_2$. The aqua colored area represents the $Nb$ layers, while the yellow area is the $Al$ layer. The process starts with the in situ deposition of a $Nb$ layer, typically $100-200\;\text{nm}$ in thickness, followed by a thin $Al$ layer of one or two nanometers. The $Al$ layer, which has excellent coverage properties of the underlying $Nb$ film is then oxidized by exposing the wafer to a well-defined $O_2$ atmosphere for a time ranging from few second to few minutes. The $O_2$ pressure and the exposure time determine the thickness of the native $Al$ oxide, normally $Al_{2}O_5$, formed. This will in turn determine the transparency of the tunnel barrier forming the JJ, so great care is taken to have a controllable and reproducible oxidation process. Subsequently, a second $Nb$ layer, again with a $100$ to $200\;\text{nm}$ thickness, is deposited on top of the previous ones, thus forming a giant and uniform JJ covering the whole. The rest of the process is devoted to the precise definition of the areas where the junctions are needed and of the relative superconducting wiring.
At first a photoresist layer is deposited, using a spinner, on the trilayer and is patterned with the areas where the junctions have to be placed. A selective $Nb$ etching process is then used to remove the uncovered top $Nb$ layer. The etching, being selective, stops at the $Al$ oxide layer. A second photoresist layer is spinned and patterned with the base electrode geometry. At this point, normally, the exposed $Al$ layer is removed using a physical etching process and immediately after the underlaying $Nb$ layer is also removed with a selective etching. After removing the resist, the base electrode pattern is defined as well as the areas where the JJ resides. 
Successive fabrication steps, not shown in Fig.~\ref{Chapter1:fig:trilayer}, consist in the deposition and patterning of an insulating layer, typically of $SiO_2$, in order to leave uncovered the top part of the JJs, and a successive $Nb$ deposition and patterning with the top electrode geometry. In this way a complete circuit with JJs can be realized. There are variants of the above process, in the way the insulation layer is formed and in the deposition of additional materials, when resistive components are needed. However, the main advantage of this way of fabricating circuits employing JJs is still the fact that the junction insulating barrier is done once and uniformly for all the junctions, thus guaranteeing a reproducible and controllable process.
As mentioned above, the main property of the Josephson tunnel barrier, namely, the maximum pair current density $J_c$, is determined by the combination of $O_2$ pressure and oxidation time in the $Al$ layer oxidation process. An empirical expression is:
\begin{equation}\label{Chapter1:J_c}
J_c \propto (P t)^n,
\end{equation}
where $J_c$ is in $A/cm^2$, $P$ in mbar, $t$ in minutes, and $n \approx 0.4 - 0.5$ ~\cite{fritzsch1998preparation, kleinsasser1995dependence}. Therefore, by properly choosing the fabrication parameters, it is possible to realize JJs having critical current densities ranging from few $A/cm^2$ to tens of thousands $A/cm^2$. Low critical current densities are typically required for the realization of the voltage standard and of ultra sensitive magnetic sensors (SQUIDs), while high critical current densities are required by (classical) digital applications.
As mentioned above, the development of quantum circuits has pushed towards the realization of JJs specifically designed for the realization of high-quality qubits. In this context, the fabrication processes must minimize sources of noise and error that can degrade qubit performance. This includes controlling material defects, reducing electromagnetic interference, and optimizing device geometries to minimize unwanted couplings. The preferred fabrication methodology adopted is based on the so called \emph{Dolan} technique~\cite{dolan1977offset}. This technique involves using shadow evaporation to create the insulating barrier in the JJ. In this process, a thin layer of insulating material is deposited onto a superconducting film through a shadow mask, creating a localized region where the insulator is present between the superconducting electrodes. Common insulating materials used include aluminum oxide $(AlO_x)$ and silicon dioxide $(SiO_2)$, chosen for their compatibility with superconducting materials and their ability to form high-quality tunnel barriers suitable for JJs. A crucial aspect of the \emph{Dolan} technique is the design of the shadow mask used during deposition. The mask defines the pattern of the insulating barrier, determining the shape and dimensions of the resulting JJs. Further discussion of the technology associated to the development of superconducting quantum circuits is beyond the scope of this chapter. The interested reader is referred to~\cite{verjauw2022path} and references therein.

\subsection{Voltage standard}

Here, we will discuss the application of JJs to the realization of the international voltage standard. Since the discovery of the Josephson effect, it was clear that JJs could be used to realize a quantum device that relates, through universal constants, the frequency of an impinging radiation and the DC voltage across the junction~\cite{Barone1982}, the so-called inverse AC Josephson effect.

\begin{figure}[t]
    \centering
    \includegraphics[width=0.75\linewidth]{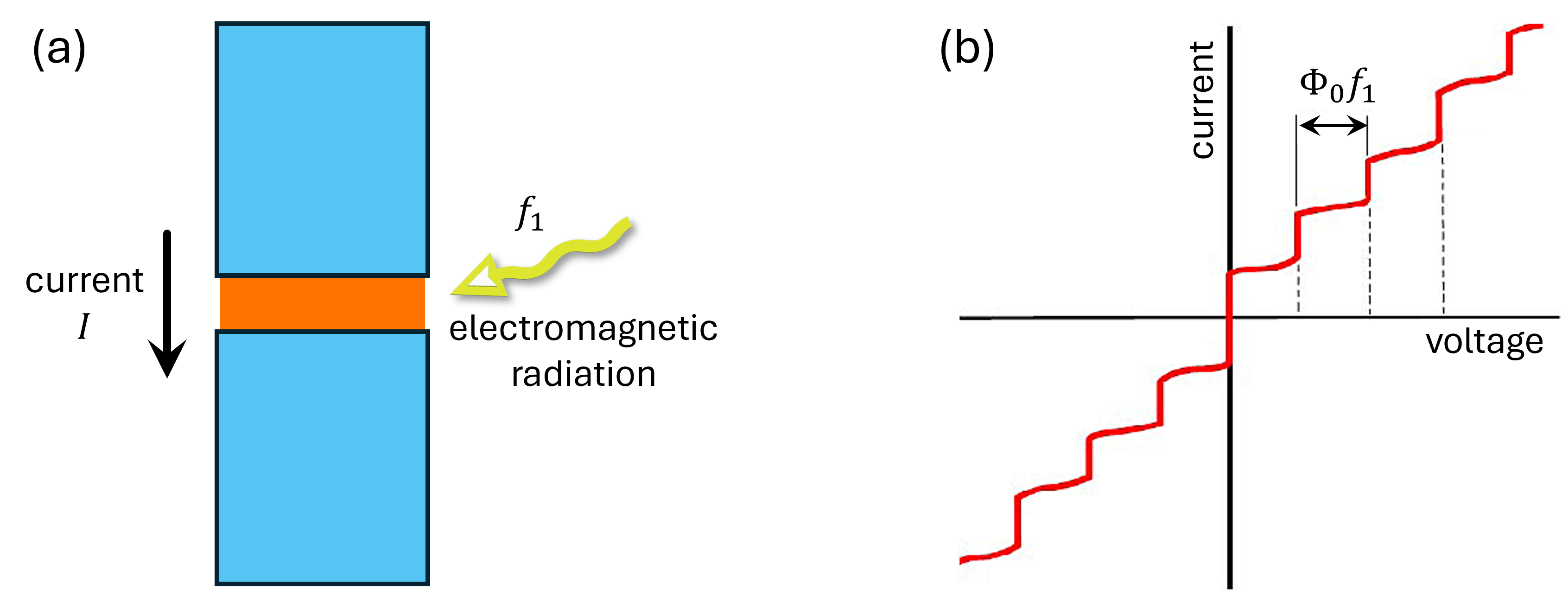}
    \caption{(a) Sketch of a JJ irradiated with an electromagnetic radiation at frequency $f_1$ producing a flowing current $I$. (b) current voltage characteristic of a JJ irradiated by an electromagnetic wave at frequency $f_{1}$. Constant voltage steps appear at voltages given by Eq.~\eqref{Chapter1:VN}.}
    \label{Chapter1:fig:RFJJ}
\end{figure}

This phenomenon has been briefly discussed in the first part of this chapter. Here, it will be recalled with some more details. A simple way to understand this situation is to consider a junction voltage biased with a DC and an RF source:
\begin{equation}\label{Chapter1:none}
V(t) = \ V_{0} + \ V_{1}\sin\left( 2\ \pi\ f_{1}\ t \right).
\end{equation}
From the second Josephson equation we have then:
\begin{equation}\label{Chapter1:none}
\varphi = \ \int_{}^{}\frac{2\ \pi}{\Phi_{0}}\ V\ dt = \ \varphi_{0} + \ \frac{2\ \pi}{\Phi_{0}}\ V_{0}\ t + \ \frac{V_{1}}{\Phi_{0}\ f_{1}}\cos\left( 2\ \pi\ f_{1}\ t \right),
\end{equation}
where
$\Phi_{0} = 2.067833831\ 10^{- 15}\ \text{Wb} = \ 2.067833831\ \mu \text{V}/\text{GHz}$
is the flux quantum, an universal constant. From the first Josephson equation, the overall
flowing current is:

\begin{eqnarray}\label{Chapter1:none}
I &=& \ I_{c}\sin\varphi = \ I_{c}\sin\left( \varphi_{0} + \ \frac{2\ \pi}{\Phi_{0}}\ V_{0}\ t + \ \frac{V_{1}}{\Phi_{0}\ f_{1}}\cos\left( 2\ \pi\ f_{1}\ t \right) \right)\ =\\\nonumber 
&=& I_{c}\sin{\left( \varphi_{0} + \frac{2\pi}{\Phi_{0}}V_{0} t \right)}\cos\left( \frac{V_{1}}{\Phi_{0} f_{1}}\cos\left( 2 \pi f_{1} t \right) \right)  + I_{c}\cos{\left( \varphi_{0} + \frac{2 \pi}{\Phi_{0}} V_{0} t \right)}\sin\left(  \frac{V_{1}}{\Phi_{0} f_{1}}\cos\left( 2 \pi f_{1} t \right) \right).
\end{eqnarray}

By using well known series expansion of $\sin(z\cos{x})$ and $\cos(z\cos{x})$ in terms of the Bessel functions, it is straightforward to derive:
\begin{equation}\label{Chapter1:none}
I = \ I_{c}\ \sum_{n = 0}^{\infty}{( - 1)^{n}\ J_{n}\left( \frac{V_{1}}{\Phi_{0}\ f_{1}} \right)\ sin\left( \varphi_{0} + \ \frac{2\ \pi}{\Phi_{0}}\ V_{0}\ t - \ 2\ \pi\ f_{1}\ t \right)}.\ 
\end{equation}

This expression represents the current as an infinite sum of
oscillations with zero average value. However, if the DC voltage assumes
one of the following values:
\begin{equation}\label{Chapter1:VN}
V_{n} = n\ \Phi_{0}\ f_{1},\qquad \text{with}\qquad n = 0, \pm 1, \pm 2,\ldots
\end{equation}
there is a net DC current flowing through the junction, appearing as a
constant-voltage current step, called \emph{Shapiro step}~\cite{Shapiro1963}, with
amplitudes given by
$\ I_{c}\ J_{n}\left( \frac{V_{1}}{\Phi_{0}\ f_{1}} \right)$, in the
current voltage characteristic curve.


In Fig.~\ref{Chapter1:fig:RFJJ}(b) such situation is shown. The relation~\eqref{Chapter1:VN} allows to connect the frequency of the impinging electromagnetic radiation, which can be known with a precision of one part in $10^{11}$, to the measured DC voltage across the junction and, through an universal
constant $\Phi_{0}$, to transfer such precision to it. The international voltage standard used at the time of the discovery
of the Josephson effect was based on the Weston cells, and had a precision
of the was one part in $10^6$. Therefore, it is clear the advantage of using the Josephson effect to define a new voltage standard.

The technological route to reach this goal was, however, difficult for two main reasons. Firstly, to obtain clean steps the voltage-bias conditions have to be fulfilled. This requires that most of the RF current flows in the junction through its capacitance and therefore, that the applied RF frequency $f_{1}$ must be greater than the junction plasma frequency $f_{p} = \ \sqrt{J_{c}/(2\ \pi\ C)}$, where \emph{C} is the junction capacitance per unit area. As typical values of $f_{p}$ are in the $GHz$ range and, in order to use available RF equipment, this implies that the junction critical current density has to be kept as small as possible. Secondly, the voltage of each step is rather small, in the few $\mu V$ range. To have a practical voltage standard, voltages of the order of $1\;V$ are necessary.

Consequently, in the last forty years arrays employing a large number of JJs, in series from the DC point of view and in parallel for the RF signal, have been developed. Current technology, based on trilayer $Nb$ fabrication with low Josephson critical current densities, has developed devices that can provide DC voltages up to $10\;V$ and also low frequency AC voltages with fundamental accuracy~\cite{tang2015application, flowers2016two}.

\subsection{SQUIDs}

\begin{figure}[t]
    \centering
    \includegraphics[width=0.75\linewidth]{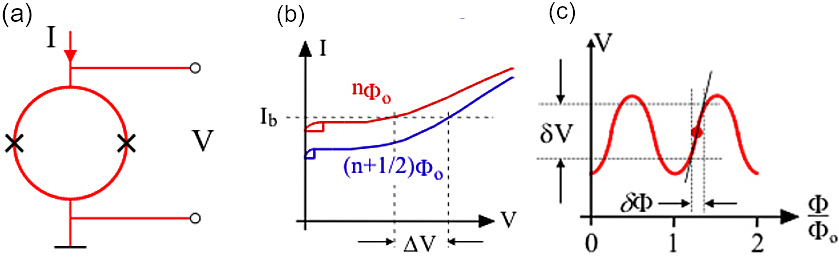}
    \caption{(a) Circuit representation of a dc-SQUID. (b) Current voltage characteristic of a dc-SQUID when an even (upper red curve), or odd (lower blue curve) number of half flux quanta are threading the loop. $I_b$ is the dc-bias current and $\Delta V$ is the voltage swing between the two flux states. (c) Dependence of the dc voltage across the SQUID on the applied magnetic flux. The red dot indicates the point of maximum responsivity.}
    \label{Chapter1:fig:SQUID}
\end{figure}

The Superconducting QUantum Interference Device (SQUID) is perhaps the most successful electronic device employing JJs. Its basic theory has been already discussed in the first part of the chapter. Here we will start from there and discuss the consequences relevant for the applications. More specifically we will concentrate on one version of the SQUID, the dc-SQUID, which is formed by a superconducting loop with to \emph{equal} junctions. In Fig.~\ref{Chapter1:fig:SQUID}(a) such situation is shown. A bias current $I$ is applied to the device and the voltage drop across it is recorded. In Fig.~\ref{Chapter1:fig:SQUID}(b) the dc current-voltage characteristic of a dc-SQUID is shown. The curve depends also on the amount of external magnetic field threading the SQUID loop, so that it oscillates from a larger current state when there is an even number of half flux quanta in the loop to a lower current state for an odd number. 
Such oscillation is periodic on the scale of one flux quantum and results in a periodic oscillation of the SQUID voltage, when it is biased at a dc current value slightly above the maximum zero voltage current, shown in Fig.~\ref{Chapter1:fig:SQUID}(c). This is the key factor for the use  of a SQUID as magnetic sensor. The red point marked in Fig.~\ref{Chapter1:fig:SQUID}(c) indeed indicates the point of maximum responsivity of the SQUID, i.e., the bias point where there is the larges variation of dc voltage in response of a small change in the applied magnetic flux.

Through an analysis of the SQUID equations, it is possible to derive an expression of its responsivity, defined as:
\begin{equation}\label{Chapter1:none}
V_{\Phi}\  \equiv \ \left. \ \frac{\delta V}{\delta\Phi} \right|_{I}\  \approx \ \ \frac{R}{L},
\end{equation}
where $R$ is the junction shunt resistance and $L$ the SQUID loop  inductance. Therefore:
\begin{equation}\label{Chapter1:none}
\delta V\  \approx \ \ \frac{R}{L}\ \delta\Phi.
\end{equation}

This result indicates that large $R$ values and small $L$ values can amplify the SQUID sensitivity. As with all sensors, the ultimate sensitivity is given by the intrinsic noise level. In SQUIDs, noise is generated by thermal fluctuations of the shunt resistors. The Nyquist voltage noise spectral density of a resistor $R$ is given by $S_{V}(f) = 4\ k_{B}T\ R$, where $k_B$ is the Boltzmann constant and $T$ the absolute temperature. This can be translated into magnetic flux spectral noise density as:
\begin{equation}\label{Chapter1:SPhi}
S_{\Phi}(f) = \frac{4\ k_{B}T\ R}{V_{\Phi}^{2}} \approx \ \frac{16\ k_{B}T\ L^{2}}{R},
\end{equation}
where the additional factor 4 on the rhs stems from more accurate numerical studies of the SQUID equations in the optimal conditions. Using realistic values for the parameters ($L = 200\ pH,\ R = 6\ \Omega,\ T = 4.2\ K)$ a value for the magnetic flux noise of about $1.2\ {\mu\Phi}_{0}/\sqrt{Hz}$ is obtained, which translates in a noise energy $\varepsilon(f) = \frac{S_{\Phi}(f)}{2L} \approx \ 10^{- 32}\frac{J}{Hz} \approx \ \ 100\ \hslash$, very close to the quantum limit.

\begin{figure}
    \centering
    \includegraphics[width=0.50\linewidth]{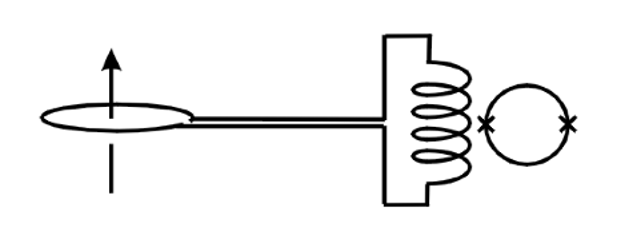}
    \caption{example of a flux transformer coupled to a dc-SQUID.}
    \label{Chapter1:fig:fluxtr}
\end{figure}

In order to translate this extreme sensitivity to magnetic flux into sensitivity to magnetic field, it is important to maximize the area of the loop that is threaded by the magnetic field to be measured. This would imply a strong increase of the loop inductance, while Eq.~\eqref{Chapter1:SPhi} calls for small values of it. A solution is the use of a superconducting flux transformer, as shown in Fig.~\ref{Chapter1:fig:fluxtr}, which, by means of its multiple windings and large pick-up area, can reach a field sensitivity of about $1\ fT/\sqrt{Hz}$.

\begin{figure}[b]
    \centering
    \includegraphics[width=0.75\linewidth]{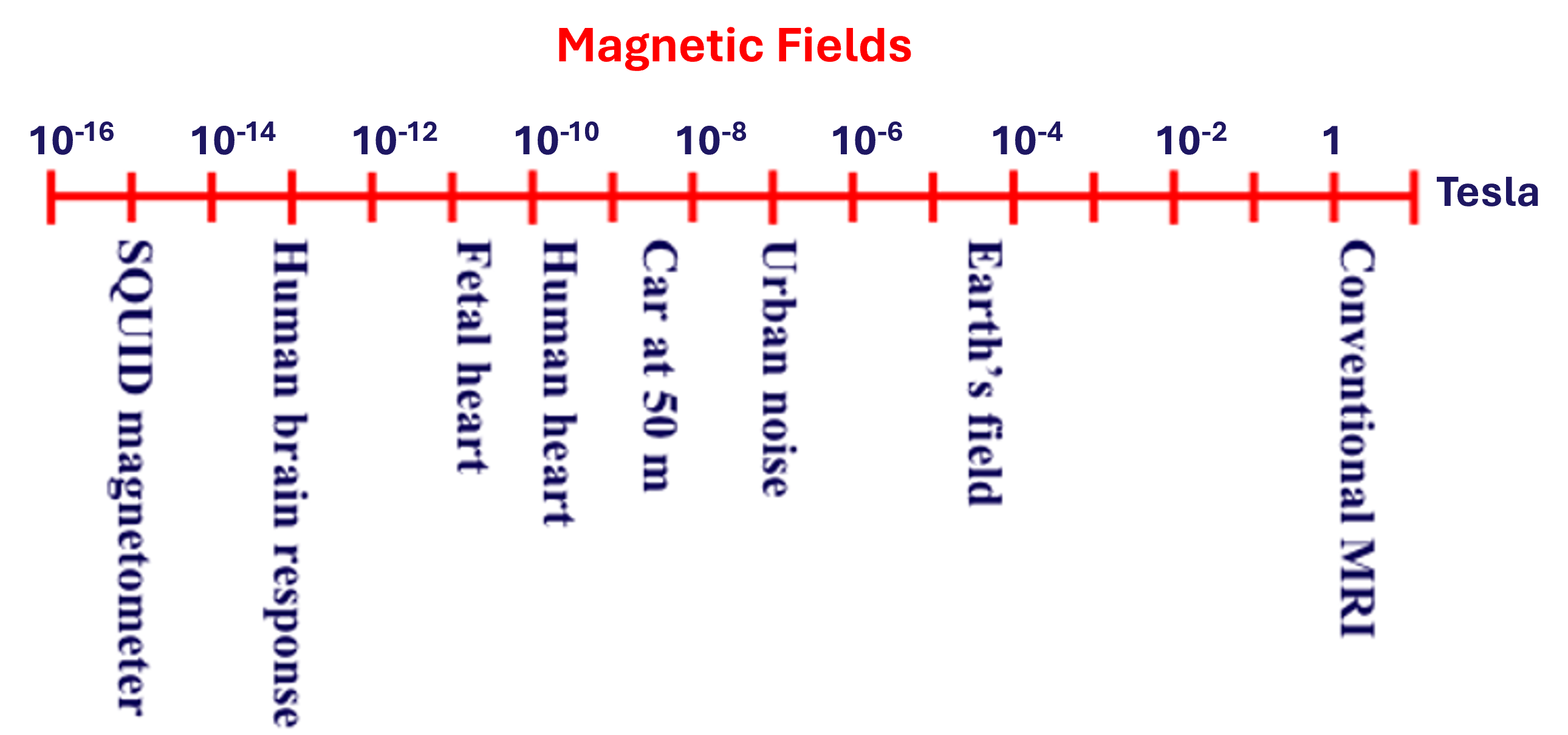}
    \caption{Typical amplitudes of magnetic fields and sensitivity level of SQUID magnetometers.}
    \label{Chapter1:fig:magnetic}
\end{figure}

Such high sensitivity paves the way to numerous applications, but also requires proper shielding from unwanted signals, that can easily saturate the SQUID response. This is achieved using a number of techniques, from the employment of passive magnetically shielded chambers to the use of active shielding, to the use of higher order (gradiometric) pickup coil configurations. Moreover, the exploitation of the nonlinear responsivity of the SQUID into a linear response device requires the use of closed loop electronics with active feedback. All these aspects are rather technical and will not be treated here.
Figure~\ref{Chapter1:fig:magnetic} shows a logarithmic scale of typical magnetic fields encountered. At the bottom there is the SQUID sensitivity.
With their extreme sensitivity, down to quantum level, SQUIDs have found in the years numerous applications in several fields. In medical imaging and diagnostics, they are at the base of Magnetoencephalography (MEG) where SQUIDs detect the weak magnetic fields generated by neural currents in the brain and help in mapping brain activity with high spatial and temporal resolution. SQUIS are also used in magnetocardiography (MCG), where they measure the magnetic fields produced by electrical currents in the heart, aiding in diagnosing cardiac conditions. In material research SQUIDS are used to investigate the magnetic properties of materials, such as superconductors, magnetic nanoparticles, and other materials at low temperatures. They also help to explore vortex dynamics in superconductors. In geophysics and environmental studies, SQUIDs play a role in geophysical prospecting. They map variations in Earth's magnetic field, assisting in mineral exploration, detecting underground water reservoirs and studying subsurface structures. In astronomy and cosmology SQUIDS are widely used as amplifiers for ultrasensitive superconducting detectors in earth and space telescopes. Finally, SQUID are integral to quantum bits (qubits) in quantum computers where their sensitivity allows precise manipulation of qubits. The interested reader can refer to the review by Kleiner \emph{et al.} and references therein~\cite{kleiner2004superconducting}. More recent applications to quantum computing can be found in~\cite{kjaergaard2020superconducting}.

\subsection{Classical digital circuits}

Since the prediction and subsequent experimental confirmation of the Josephson effect, it was clear that JJs could be used as switching elements. The presence of two clearly distinguishable states, i.e., zero and finite, although small, voltage, and a sub ns response time for the passage between these states, made them ideal candidates as binary logic elements. Already in 1967 the IBM Corporation started a Josephson computer project that, in few years and based on the technologies then available, was able to develop a prototype processor running at the astonishing clock of 1 GHz~\cite{anacker1980Josephson}. The history of Josephson computing has had several ups and down, due to internal reasons: new technologies and new designs, and to external ones: the everlasting competition with semiconductors~\cite{dorojevets2004road}. The Moore's law is universally known and has led the performances of semiconducting digital circuits through an exponential growth in the last almost sixty years. What was considered an enormous advantage in terms of speed and packaging density back in 1980 has now been largely overtaken by modern semiconducting processors. However, JJs have demonstrated a surprising vitality, thanks to new ways to exploit its intrinsic quantum nature.

\begin{figure}
    \centering
    \includegraphics[width=0.75\linewidth]{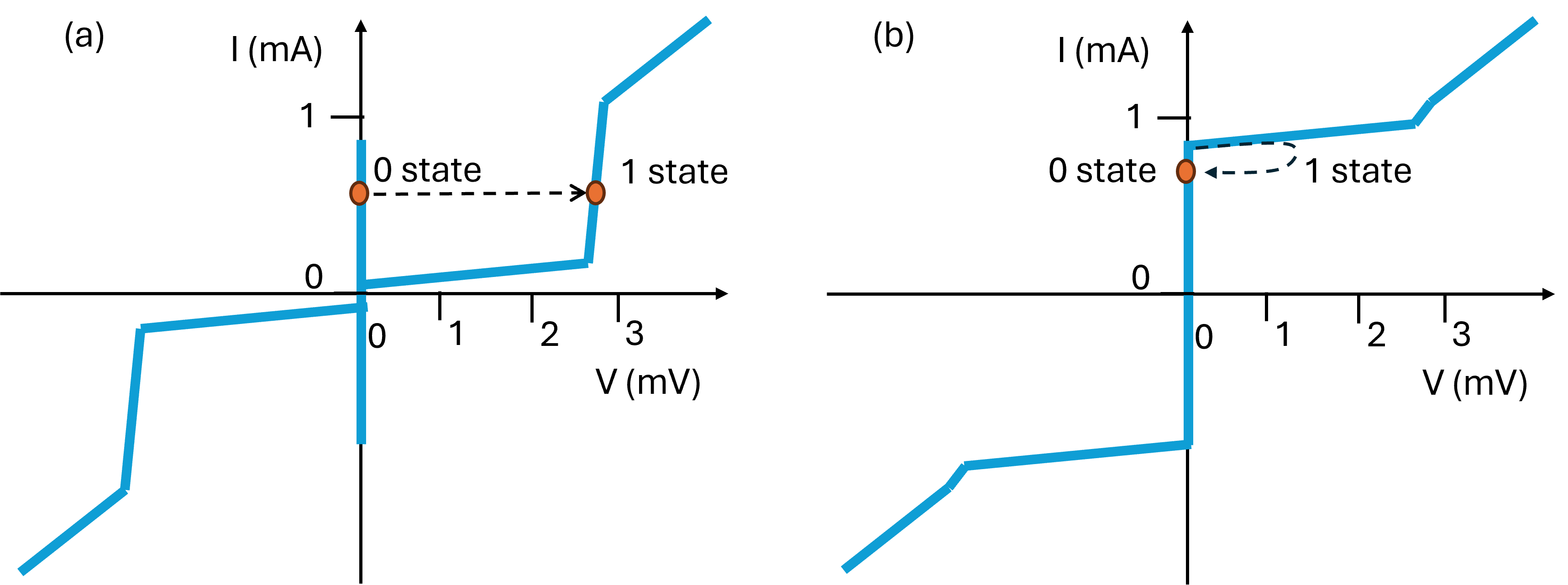}
    \caption{Current-voltage characteristics of a hysteretic (a) and a non-hysteretic one (b). The dots represent the stable bias states.}
    \label{Chapter1:fig:IVdig}
\end{figure}

Figure~\ref{Chapter1:fig:IVdig} shows the current-voltage characteristics of two different types of junctions used to realize digital devices: a hysteretic junction and a non-hysteretic one. In the first case the binary state is associated to two stable bias points, shown as orange dots in the figure. The binary ``0" is associated the zero-voltage state and the binary ``1" to the finite voltage state (about 2.5 mV for a $Nb$ junction). The transition between the states is achieved by a positive current pulse, for the $0\longrightarrow1$ case, and a negative current pulse for the $1\longleftarrow0$ case. It is worth noting that the curves in Fig.~\ref{Chapter1:fig:IVdig} represent the average voltage measured across the junction, as the Josephson oscillations, being of the order of few GHz, cannot be recorded by ordinary electronic equipment. The time it takes for the junction to stabilize in the new bias point after a transition determines the maximum speed of operation of the logic device. For typical $Nb$-based junctions, its dynamics limits such speed to a rate of about 1 GHz. This sets the maximum operating clock frequency of the device. Such value, which was at least one order of magnitude higher than of that of the large IBM mainframes in the 1980s, is nowadays easily overcome by any portable computing device. This, together with other material related reliability problems lead IBM to abandon the Josephson computer project in 1983. Meanwhile other institutions, in US and in Japan kept working on Josephson digital circuits development, also thanks to the advent of the trilayer $Nb$ technology, discussed above. In 1991 the group of K. Likharev proposed a new mechanism for logic operations using JJs~\cite{likharev1991rsfq}, which was not affected by the speed limitation of the voltage state mechanism. The new logic operation was called Rapid Single Flux Quantum (RSFQ) and has been demonstrated to operate up to frequencies of few hundreds of GHz, an unreachable value for semiconducting electronics.

The operation of RSFQ logic is quite different from the voltage state one. It is a clocked type of logic, that is it requires a reference train of pulses (the clock). The Josephson gate, in certain conditions, can generate a Single Flux Quantum (SFQ) pulse and the logical $1$ and $0$ states are encoded as the presence or absence of an SFQ pulse between two clock pulses. The junction is resistively shunted, in order to show a non-hysteretic current voltage characteristic (see section 1 of this chapter), inserted in a superconductive loop and dc biased near it critical current value (the orange dot in Fig.~\ref{Chapter1:fig:IVdig}b). Upon arrival of a current pulse, the junction switches momentarily to the voltage state, but, being the current-voltage curve single valued, it has to self-reset to its initial state (see dashed curve in Fig.~\ref{Chapter1:fig:IVdig}b). This occurs in few ps and corresponds to a $2 \pi$ phase jump across the junction. From the second Josephson equation a phase jump corresponds to a voltage pulse with a quantized area corresponding to exactly one flux quantum, hence
the name SFQ pulse.
\begin{equation}\label{Chapter1:none}
\int_{V(\varphi = \varphi_{0})}^{V(\varphi = \varphi_{0} + 2\ \pi)}{V(t)\ dt} = \ \int_{\varphi_{0}}^{\varphi_{0} + 2\pi}\frac{\Phi_{0}}{2\ \pi}\ \frac{d\varphi}{dt}\ dt = \ \Phi_{0} = 2.07\ mV\ ps\ 
\end{equation}

Such flux quantum can be trapped in the superconducting loop or moved, through JJs, from one loop to another as millivolt picosecond SFQ pulses. In this way this basic bit of information can be manipulated and perform logic operations in a few picoseconds. Such short pulses require impedance matched microstrip lines to propagate over adjacent gates, which represents a design challenge, only recently addressed by the semiconductor industry.

A detailed description of the operation of Josephson logic gates is beyond the scope of this chapter. Many details regarding both types of logics can be found in standard textbooks~\cite{Barone1982,Likharev1986,van1981principles}, while a more recent survey on superconducting computing technology can be found in~\cite{huang2022survey}.

Overall, the superconducting digital technology has demonstrated a good capability to withstand the overwhelming increase of performances exhibited by semiconductive circuits. From one side, the level of integration reached in modern processors is unreachable by superconductors for many reasons, also economic. However, superconductive circuits outperforms in terms of speed and low dissipated power and, in perspective, can effectively team with semiconductors in applications where ultra-fast signal processing is needed while leaving slower tasks, like storage, to the ultra-dense semiconductive circuits.


\section{Towards quantum computing with superconducting qubits}
\label{Chapter1:sec:quantum-computing}

In this Section the basics of quantum computing will be described. Starting from the Bloch sphere representation for the qubit, we will introduce the quantum gates and their operations, to conclude with an overview on the quantum supremacy and state of the art quantum computers.

\subsection{Bloch sphere representation}

\begin{figure*}[ht]
\centering
    \includegraphics[width=5cm]{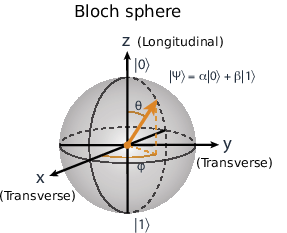}
    \caption{
     Bloch sphere representation of the quantum state $\lvert \psi \rangle = \alpha \lvert 0 \rangle + \beta \lvert 1 \rangle$. The qubit quantization axis -- the $z$ axis -- is \emph{longitudinal} in the qubit frame, corresponding to $\sigma_z$ terms in the qubit Hamiltonian. The $x$-$y$ plane
    is \emph{transverse} in the qubit frame, corresponding to $\sigma_x$ and $\sigma_y$ terms in the qubit Hamiltonian.
   }
\label{Chapter1:fig:Bloch-sphere}
\end{figure*}

As we have seen in the previous Chapter, the superconducting qubit is a type of qubit that utilizes superconducting circuits to exploit the unique properties of superconductors for quantum computation and its basic building block is a JJ.
Generically the qubit is a two-level system that can be represented as a superposition of two states ({\it quantum state}). The \emph{Bloch sphere} is a unit sphere used to represent the quantum state of the qubit. In Fig.~\ref{Chapter1:fig:Bloch-sphere} the Bloch sphere is shown with a \emph{Bloch vector} representing the state $\lvert \psi\rangle = \alpha \lvert 0\rangle + \beta \lvert 1\rangle$. By convention, the north pole represents state $\lvert 0 \rangle$ and the south pole state $\lvert 1 \rangle$. For pure quantum states such as $\lvert \psi \rangle$, the Bloch vector is of unit length, $\lvert \alpha \rvert^2+\lvert \beta \rvert^2=1$, connecting the center of the sphere to any point on the surface.

The $z$-axis connects the north and south poles. It is called the \emph{longitudinal axis} and represents the \emph{qubit quantization axis} for the states $\lvert 0 \rangle$ and $\lvert 1 \rangle$ in the qubit eigenbasis. Following our convention, state $\lvert 0 \rangle$ at the north pole is associated with $+1$, and state $\lvert 1 \rangle$ (the south pole) with $-1$. In turn, the $x$-$y$ plane is the \emph{transverse plane} and $x$ and $y$ are called \emph{transverse axes}. In a parametric representation, the unit Bloch vector $\vec{a}=(\sin \theta \cos \phi, \sin \theta \sin \phi, \cos \theta)$ is represented using the polar angle
$0\leq \theta \leq \pi$ and the azimuthal angle $0 \leq \phi < 2\pi$, as illustrated in Fig.~\ref{Chapter1:fig:Bloch-sphere}.  We can similarly represent the quantum state using the angles $\theta$ and $\phi$,
\begin{equation}
    \lvert \psi \rangle = \alpha \lvert 0 \rangle + \beta \lvert 1 \rangle = \cos\frac{\theta}{2} \lvert 0 \rangle + e^{i \phi}\sin \frac{\theta}{2} \lvert 1 \rangle.
    \label{Chapter1:Eq:Bloch-vector-stationary}
\end{equation}
The Bloch vector is stationary on the Bloch sphere in the \emph{rotating frame picture}. If state $\lvert 1 \rangle$ has a higher energy than state $\lvert 0 \rangle$ (as it generally does in superconducting qubits), then in a stationary frame, the Bloch vector would precess around the $z$-axis at the qubit frequency $(E_1-E_0)/\hbar$. Without loss of generality (and much easier to visualize), we instead \emph{choose} to view the Bloch sphere in a reference frame where the $x$ and $y$-axes also rotate around the $z$-axis at the qubit frequency. In this \emph{rotating frame}, the Bloch vector appears stationary as written in Eq.~(~\ref{Chapter1:Eq:Bloch-vector-stationary}).

For completeness, we note that the density matrix $\rho=\lvert \psi \rangle \langle \psi \vert$ for a pure state $\lvert \psi \rangle$ is equivalently
\begin{align}
    \rho = \frac{1}{2}(I + \vec{a} \cdot \vec{\sigma})
    &= \frac{1}{2}
    \left(
        \begin{matrix}
            1+\cos \theta & e^{-i\phi} \sin \theta \\
            e^{i\phi} \sin \theta  & 1+ \sin \theta
        \end{matrix}
    \right) \\
     &= \left(
        \begin{matrix}
            \cos^2\frac{\theta}{2} & e^{-i\phi}\cos\frac{\theta}{2} \sin\frac{\theta}{2} \\
            e^{i\phi}\cos\frac{\theta}{2} \sin\frac{\theta}{2}  & \sin^2\frac{\theta}{2}
        \end{matrix}
    \right) \\
    &=\left(
        \begin{matrix}
            |\alpha|^2 & \alpha \beta^*  \\
            \alpha^* \beta  & |\beta|^2
        \end{matrix}
    \right),
    \label{Chapter1:eq:rho}
\end{align}
where $I$ is the identity matrix, and $\vec{\sigma} = [\sigma_x, \sigma_y, \sigma_z]$ is a vector of Pauli matrices.
If the Bloch vector $\vec{a}$ is a unit vector, then $\rho$ represents a pure state, $\psi$, and $\mathrm{Tr}(\rho^2)=1$. 
More generally, the Bloch sphere can be used to represent \emph{mixed states}, for which $\lvert \vec{a} \rvert < 1$; in this case, the Bloch vector terminates at points \emph{inside} the unit sphere, and $0\leq \mathrm{Tr}(\rho^2)<1$. To summarize, the surface of the unit sphere represents pure states, and its interior represents mixed states~\cite{Nielsen2011}.

\subsection{\label{Chapter1:sec:QubitControl}Qubit control}
In this section, we will introduce how superconducting qubits are manipulated to implement quantum algorithms.
The transmon-like variant of superconducting qubits has emerged as the most extensively utilized modality for implementing quantum programs. Consequently, the discussion in this section will concentrate on methodologies tailored for transmons. It is noteworthy, however, that the techniques presented herein can be extended to all categories of superconducting qubits.

Thus, we make a brief review on the gates used in classical computing as well as quantum computing, and the concept of universality. Subsequently we discuss the most common technique of driving single qubit gates via a capacitive coupling of a microwave line, coupled to the qubit. We introduce the notion of ``virtual $\Z{}$ gates'' and ``DRAG'' pulsing. In the latter part of this section, we review some of the most common implementations of two-qubit gates in both tunable and fixed-frequenzy transmon qubits. The single-qubit and two-qubit operations together form the basis of many of the medium-scale superconducting quantum processors that exist today.

Throughout this section, we write everything in the computational basis $\{|0\rangle,|1\rangle\}$ where $|0\rangle$ is the $+1$ eigenstate of $\sigma_z$ and $|1\rangle$ is the $-1$ eigenstate. We use capitalized serif-fonts to indicate the rotation operator of a qubit state, e.g. rotations around the $x$-axis by an angle $\theta$ is written as
\begin{equation}
 \textsf{X}_{\theta} = R_X(\theta) = e^{-i\frac{\theta}{2}\sigma_x} = \cos(\theta/2)\Id-i\sin(\theta/2) \sigma_x
\end{equation}
and we use the shorthand notation `$\X{}$' for a full $\pi$ rotation about the $x$ axis (and similarly for $\Y{} := \Y{\pi}$ and $\Z{} := \Z{\pi}$).

\subsection{\label{Chapter1:sec:ClassicalGatesInQC}Boolean logic gates in classical computers}

\begin{figure}[!h]
\centering
\includegraphics[width=8cm]{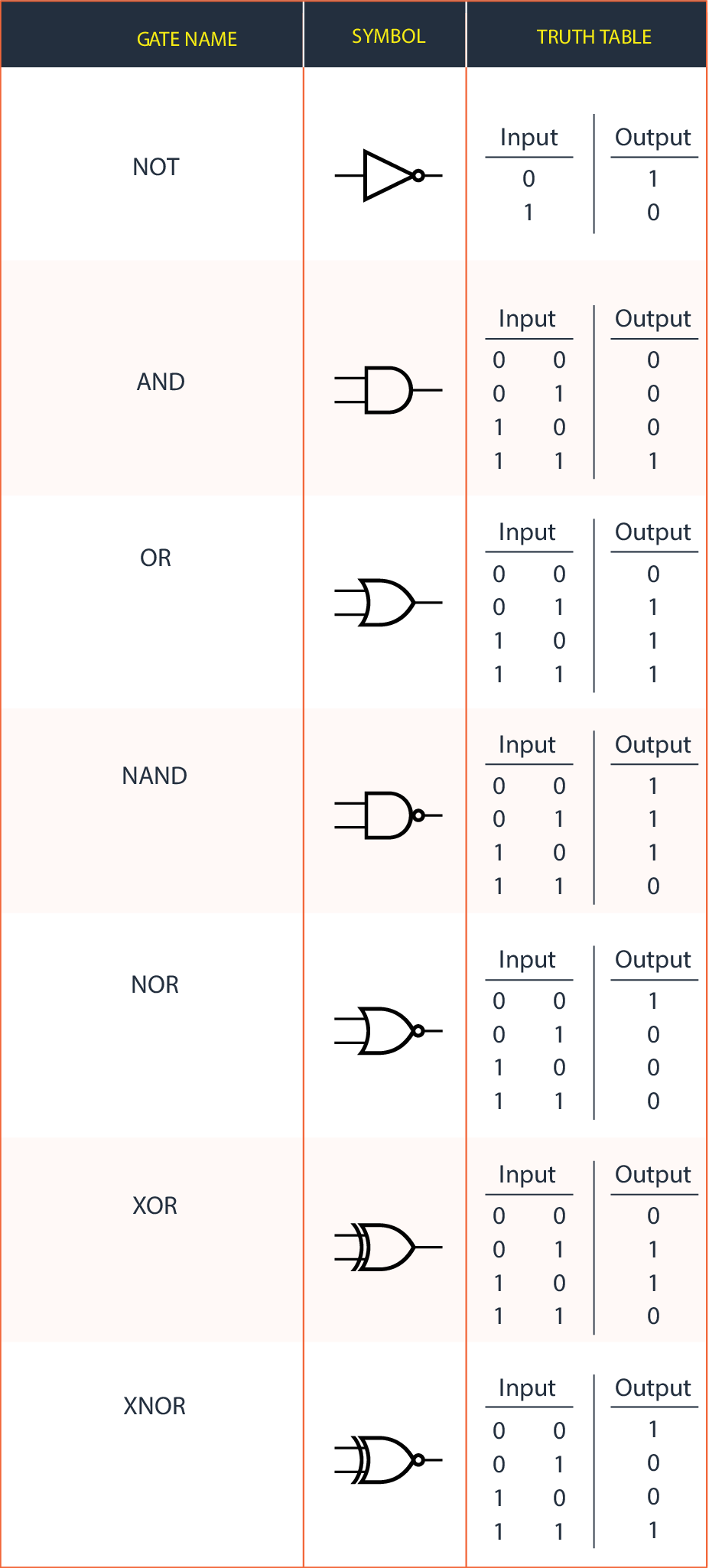}
\caption{Classical single-bit and two-bit boolean logic gates. For each gate, the name, a short description, circuit representation, and input/output truth tables are presented. The numerical values in the truth table correspond to the classical bit values $0$ and $1$. Modified from \cite{krantz2019quantum}, \textcopyright{} The Authors, MIT xPro Quantum Curriculum, \href{https://learn-xpro.mit.edu/quantum-computing}{https://learn-xpro.mit.edu/quantum-computing}, 2018. All rights reserved.}
\label{Chapter1:fig:Classical-logic-gates}
\end{figure}

Boolean logic can be implemented on classical computers using a small set of single-bit and two-bit gates. In Fig.~\ref{Chapter1:fig:Classical-logic-gates} several classical logic gates are shown along with their truth tables. In classical boolean logic, bits can take on one of two values: state $0$ or state $1$. The state $0$ represents logical \textsf{FALSE}, and state $1$ represents logical \textsf{TRUE}.

Beyond the trivial \emph{identity operation}, which simply gives a boolean bit unchanged, the only other possible single-bit boolean logic gate is the \textsf{NOT} gate. As shown in Fig.~\ref{Chapter1:fig:Classical-logic-gates}, the \textsf{NOT} gate flips the bit: $0 \rightarrow 1$ and $1 \rightarrow 0$. This gate is \emph{reversible}, because it is trivial to determine the input bit value given the output bit values. As we will see, for two-bit gates, this is not the case.

\begin{figure*}[!h]
\centering
\includegraphics[width=8cm]{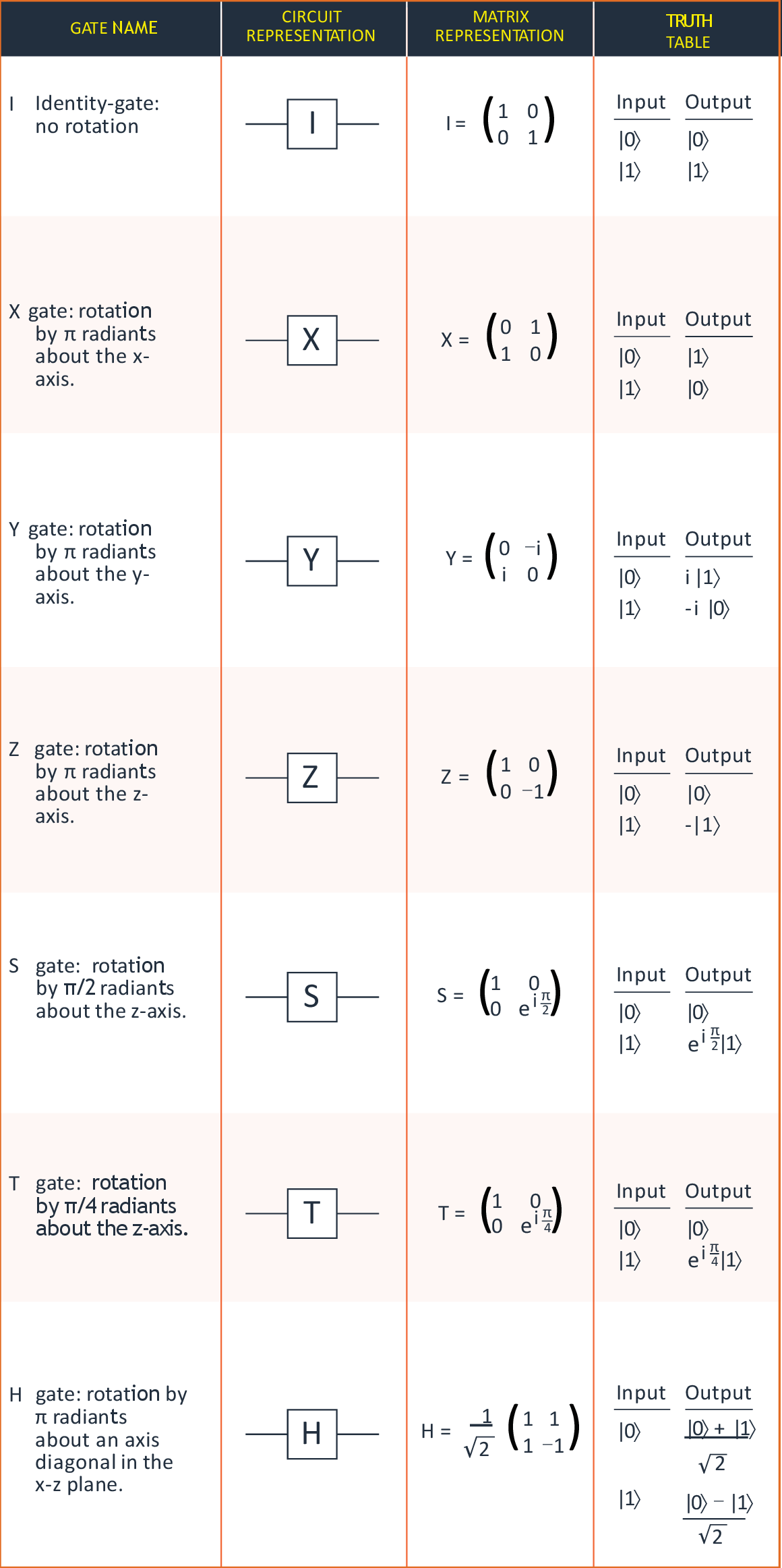}
\caption{Quantum single-qubit gates. For each gate, the name, circuit representation, matrix representation, input/output truth tables are presented. Matrices are defined in the basis spanned by the state vectors $\vert 0 \rangle \equiv [ 1 \; 0 ]^T$ and $\vert 1 \rangle \equiv [ 0 \; 1 ]^T$. The numerical values in the truth table correspond to the quantum states $\vert 0 \rangle$ and $\vert 1 \rangle$. Modified from \cite{krantz2019quantum}, \textcopyright{} The Authors, MIT xPro Quantum Curriculum, \href{https://learn-xpro.mit.edu/quantum-computing}{https://learn-xpro.mit.edu/quantum-computing}, 2018. All rights reserved.}
\label{Chapter1:fig:Quantum-logic-gates-single}
\end{figure*}

There are several two-bit gates shown in Fig.~\ref{Chapter1:fig:Classical-logic-gates}. A two-bit gate takes two bits as inputs, and it gives as an output the result of a boolean operation. One common example is the \textsf{AND} gate, for which the output is $1$ if and only if both inputs are $1$; otherwise, the output is $0$. The \textsf{AND} gate, and the other two-bit gates shown in Fig.~\ref{Chapter1:fig:Classical-logic-gates}, are all examples of \emph{irreversible} gates; that is, the input bit values cannot be inferred from the output values. For example, for the \textsf{AND} gate, an output of logical $1$ uniquely identifies the input $11$, but an output of $0$ could be associated with $00$, $01$, or $10$. Once the operation is performed, in general, it cannot be undone and the input information is lost.
\noindent There are several variants of two-bit gates, including,
\begin{itemize}
    \item \textsf{AND} and \textsf{OR};
    \item \textsf{NAND} (a combination of \textsf{NOT} and \textsf{AND}) and \textsf{NOR} (a combination of \textsf{NOT} and \textsf{OR});
    \item \textsf{XOR} (exclusive \textsf{OR}) and \textsf{NXOR} (\textsf{NOT XOR}).
\end{itemize}
The \textsf{XOR} gate is also known \emph{parity} gate. That is, it returns a logical $0$ if the two inputs are the same values (i.e., they have the same parity), and it returns a logical $1$ if the two inputs have different values (i.e., different parity). Still, the \textsf{XOR} and \textsf{NXOR} gates are not reversible, because knowledge of the output does not allow one to uniquely identify the input bit values.

\emph{Universality}, in the context of quantum computing, pertains to the capability of executing any boolean logic algorithm using a small set of single-bit and two-bit gates.
A universal gate set can in principle transform any state to any other state in the state space represented by the classical bits. The set of gates that enable universal computation is not unique, and may be represented by a small set of gates. For example, the \textsf{NOT} gate and the \textsf{AND} gate together form a universal gate set. Similarly, the \textsf{NAND} gate itself is universal, as is the \textsf{NOR} gate. The efficiency of implementation depends on the choice of the gate set.

\subsection{Quantum logic gates used in quantum computers}
\label{Chapter1:sec:QuantumGatesInQC}

Quantum logic can similarly be performed by a small set of single-qubit and two-qubit gates. Qubits can of course assume the classical states $\vert 0 \rangle$ and $\vert 1 \rangle$, at the north pole and south pole of the Bloch sphere, but they can also assume arbitrary superpositions $\alpha \vert 0 \rangle + \beta \vert 1 \rangle$, corresponding to any other position on the sphere.

Single-qubit operations translate an arbitrary quantum state from one point on the Bloch sphere to another point by rotating the Bloch vector (spin) a certain angle about a particular axis. As shown in Fig.~\ref{Chapter1:fig:Quantum-logic-gates-single}, there are several single-qubit operations, each represented by a matrix that describes the quantum operation in the computational basis represented by the eigenvectors of the $\sigma_z$ operator, i.e. $\vert 0 \rangle \equiv [ 1 \; 0 ]^T$ and $\vert 1 \rangle \equiv [ 0 \; 1 ]^T$.

For example, the \emph{identity gate} performs no rotation on the state of the qubit. This is represented by a two-by-two identity matrix. The $\X{}$-gate performs a $\pi$ rotation about the $x$ axis. Similarly, the $\Y{}$-gate and $\Z{}$-gate perform a $\pi$ rotation about the $y$ axis and $z$ axis, respectively. The $\textsf{S}$-gate performs a $\pi/2$ rotation about the $z$ axis, and the $\text{T}$-gate performs a rotation of $\pi/4$ about the $z$ axis. The Hadamard gate $\textsf{H}$ is also a common single-qubit gate that performs a $\pi$ rotation around an axis diagonal in the $x$-$z$ plane, see Fig.~\ref{Chapter1:fig:Quantum-logic-gates-single}.

Two-qubit quantum-logic gates are generally \emph{conditional} gates and take two qubits as inputs. Typically, the first qubit is the \emph{control} qubit, and the second is the \emph{target} qubit. A unitary operator is applied to the target qubit, dependent on the state of the control qubit. The two common examples 
are the controlled NOT (\CNOT{}-
gate) and controlled phase (\textsf{CZ} or \CPHASE{} gate). The \CNOT{}-gate flips the state of the target qubit conditioned on the control qubit being in state $\vert 1 \rangle$. The \CPHASE-gate applies a $\Z{}$ gate to the target qubit, conditioned on the control qubit being in state $\vert 1 \rangle$. As we will shown later, the $i\textsf{SWAP}$ gate -- another two-qubit gate -- can be built from the \textsf{CNOT}-gate and single-qubit gates.\\
The unitary operator of the $\CNOT{}$ gate can be written in a useful way, highlighting that it applies an $\X{}$ depending on the state of the control qubit.
\begin{equation}
U_\CNOT{} =
\begin{bmatrix}
1 & 0 & 0 & 0 \\
0 & 1 & 0 & 0 \\
0 & 0 & 0 & 1 \\
0 & 0 & 1 & 0 \\
\end{bmatrix} = |0\rangle\langle  0| \otimes \Id + |1\rangle\langle 1| \otimes \textsf{X} \label{Chapter1:eq:UCNOT}
\end{equation}
and similarly for the $\CPHASE{}$ gate,
\begin{equation}
U_\CPHASE{} =
\begin{bmatrix}
1 & 0 & 0 & 0 \\
0 & 1 & 0 & 0 \\
0 & 0 & 1 & 0 \\
0 & 0 & 0 & -1 \\
\end{bmatrix} = |0\rangle\langle  0| \otimes \Id + |1\rangle\langle 1| \otimes \textsf{Z}\ .
\label{Chapter1:eq:UCPHASE}
\end{equation}
Comparing the last equality above with the unitary for the \CNOT{} [~\ref{Chapter1:eq:UCNOT}], it is clear that the two gates are closely related. Indeed, a \CNOT{} can be generated from a \CPHASE{} by applying two Hadamard gates,
\begin{equation}
U_\CNOT = (\Id \otimes \textsf{H})U_\CPHASE(\Id \otimes \textsf{H}),
\end{equation}
since $\textsf{H}\textsf{Z}\textsf{H} = \textsf{X}$. Due to the form of~\ref{Chapter1:eq:UCPHASE}, the \CPHASE{} gate is also denoted the \CZ{} gate, since it applies a controlled $\Z{}$ operator, by analogy with \CNOT{} (a controlled application of $\X{}$ operator).

Some two-qubit gates such as \CNOT{} and \CPHASE{} are also called \emph{entangling gates}, because they can take product states as inputs and output entangled states. They are thus an indispensable component of a universal gate set for quantum logic. For example, consider two qubits $A$ and $B$ in the following state:
\begin{equation}
    \vert \psi \rangle = \frac{1}{\sqrt{2}} \left(\vert 0 \rangle + \vert 1 \rangle\right)_A \vert 0 \rangle_B.
\end{equation}
If we perform a \CNOT{} gate, $U_{\CNOT}$, on this state, with qubit A the control qubit, and qubit B the target qubit, the resulting state is:
\begin{equation}
    U_{\CNOT} \vert \psi \rangle = \frac{1}{\sqrt{2}} \left(\vert 0 \rangle_A \vert 0 \rangle_B + \vert 1 \rangle_A \vert 1 \rangle_B \right) \neq ( \ldots)_A ( \ldots)_B,
\end{equation}
which is a state that cannot be factored into an isolated qubit-A component and a qubit-B component. This is one of the two-qubit entangled \emph{Bell states}, a manifestly quantum mechanical state.

A universal set of single-qubit and two-qubit gates is sufficient to implement arbitrary quantum logic. This means that this gate set can in principle reach \emph{any} state in the multi-qubit state-space. How efficiently this is done depends on the choice of quantum gates that comprise the gate set. We also note that each of the single-qubit and two-qubit gates is \emph{reversible}, that is, given the output state, one can uniquely determine the input state. As we discuss further, this distinction between classical and quantum gates arises, because quantum gates are based on \emph{unitary} operations $U$. If a unitary operation $U$ is a particular gate applied to a qubit, then its hermitian conjugate $U^{\dagger}$ can be applied to recover the original state, since $U^{\dagger}U=I$ resolves an identity operation.


\subsection{Classical versus quantum gates}
\label{Chapter1:sec:GatesInQC}
\begin{figure}[!ht]
\centering
\includegraphics[width=8cm]{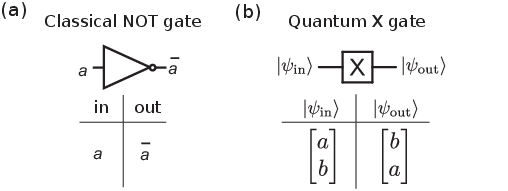}
\caption{Comparison of the classical  (\textsf{NOT}) gate and quantum (\textsf{X}) gate. \textbf{(a)} The classical \textsf{NOT} gate that inverts the state of a classical bit. \textbf{(b)} The quantum \textsf{X} gate, which flips the coefficients of the two components of the quantum bit.}
\label{Chapter1:fig:ClassicalVQuantum_circuits}
\end{figure}

The gate-sequences used to represent quantum algorithms have certain similarities to those used in classical computing, with a few striking differences. As an example, we consider first the classical \textsf{NOT} gate (discussed previously), and the related quantum circuit version, shown in Fig.~\ref{Chapter1:fig:ClassicalVQuantum_circuits}.

While the classic bit-flip gate inverts any input state, the quantum bit-flip does not in general produce the antipodal vector, but rather exchanges the prefactors of the wavefunction. The \textsf{X} operator is sometimes referred to as `the quantum NOT', but we note that \textsf{X} only acts similar to the classical \textsf{NOT} gate in the case of classical data stored in the quantum bit, i.e. $\textsf{X}|g\rangle = |\bar g\rangle$ for $g\in\{0,1\}$.

As briefly mentioned in Sec.~\ref{Chapter1:sec:QuantumGatesInQC}, \emph{all} quantum gates are \emph{reversible}, due to the underlying unitary nature of the operators implementing the logical operations. Certain other processes used in quantum information processing, however, are irreversible. Namely, measurements and energy loss to the environment. These processes are modeled by the \emph{amplitude damping} or \emph{phase damping} operator, but we refer the interested reader, e.g., to Ref.~\cite{Nielsen2011}. Let us finally note that quantum circuits are written left-to-right (in order of application), while the calculation of the result of a gate-sequences, e.g the circuit
\begin{equation}
\Qcircuit @C=1em @R=2em{
 \lstick{|\psi_\text{in}\rangle} & \gate{U_0} & \gate{U_1} & \qw & \cdots & & \gate{U_n} & \rstick{|\psi_\text{out}\rangle} \qw\\
}
\label{Chapter1:eq:simplecirc}
\end{equation}
is performed right-to-left, i.e.
\begin{equation}
|\psi_\text{out}\rangle = U_n \cdots U_1U_0 |\psi_\text{in}\rangle.
\end{equation}
As discussed in Sec.~\ref{Chapter1:sec:ClassicalGatesInQC}, the \textsf{NOR} and \textsf{NAND} gates are each individually universal gates for classical computing. Since both of these gates have no direct quantum analogue (because they are not reversible), it is natural to ask which gates \emph{are} needed to build a universal quantum computer. It turns out that the ability to rotate about arbitrary axes on the Bloch-sphere (i.e., a complete single-qubit gate set), supplemented with any entangling 2-qubit operation will suffice for universality~\cite{Barenco1995,Nielsen2011}. By using what is known as the `Krauss-Cirac decomposition', any two-qubit gate can be decomposed into a series of \CNOT{} operations~\cite{Nielsen2011}.

\subsubsection{Minimum gate sets}
\label{Chapter1:sec:gatesetgatesynth}
A universal quantum gate set is
\begin{equation}
\mathcal G_0 = \{\X{\theta},\Y{\theta},\Z{\theta},\text{Ph}_\theta,\CNOT{}\}
\end{equation}
where Ph$_\theta = e^{i\theta} 1$ applies an overall phase $\theta$ to a single qubit. For completeness we mention another universal gate set, which is of particular interest from a theoretical perspective, namely
\begin{equation}
\mathcal G_1 = \{\textsf{H}, \S, \T, \CNOT{}\}.
\end{equation}
Let us note that the restriction to a discrete gate set still gives rise to universality. This fact relies on using the so-called Solovay-Kitaev~\cite{Kitaev1997,Dawson2006} theorem, which states that any other single-qubit gate can be approximated to an error $\epsilon$ using only $\mathcal O(\log^g (1/\epsilon))$ (where $g>0$) single-qubit gates from $\mathcal G_1$. 
The gate-set $\mathcal{G}_1$ is typically referred to as the `Clifford + $T$' set, 
where $\textsf{H}$, $\S$ and $\CNOT{}$ are all Clifford gates.

Each quantum computing architecture will have certain gates that are simpler to implement at the hardware level than others (these are referred to as 'native' gates). These are typically the gates for which the Hamiltonian governing the gate-implementation gives rise to a unitary propagator that corresponds to the gate itself. Regardless of which gates are natively available, as long as one has a complete gate set, one can use the Solovay-Kitaev theorem to synthesize any other set efficiently. In general one wants to keep the overall number of time steps in which gates are applied (known as \emph{depth} of a quantum circuit) as low as possible, and one wants to use as many of the native gates as possible. Moreover, running a quantum algorithm also depends on the qubit connectivity of the device. The process of designing a quantum gate sequence that efficiently implements a specific algorithm is known as \emph{gate synthesis} and \emph{gate compilation}, respectively. A full discussion of this research line can be found in Refs.~\cite{Chong2017,Campbell2017} and references therein as a starting point. As a concrete (and trivial) example of how gate identities can be used, in~\ref{Chapter1:eq:Hfrompulses} we illustrate how the Hadamard gate from $\mathcal G_1$ can be generated by two single-qubit gates (from $\mathcal G_0$) and an overall phase gate,

\begin{align}
\textsf{H} &= \text{Ph}_{\frac{\pi}{2}}\Y{\frac{\pi}{2}} \Z{\pi} = i\frac{1}{\sqrt{2}} \begin{bmatrix}
1 & -1\\
1 & 1
\end{bmatrix}
\begin{bmatrix}
-i & 0\\
0 & i
\end{bmatrix}=
\frac{1}{\sqrt{2}}
\begin{bmatrix}
1 & 1\\
1 & -1
\end{bmatrix}
\label{Chapter1:eq:Hfrompulses}
\end{align}
The gates $\X{\theta}$, $\Y{\theta}$ and $\Z{\theta}$ are all natively available in a superconducting quantum processor.
The question on how single qubit rotations and two-qubit operations are implemented in transmon-based superconducting quantum processors have already been addressed in the previous Chapter.

\subsection{Further developments for superconducting qubits}
As discussed up to now, the planar superconducting qubits represents a promising platform for realizing medium scale quantum processors. While we have focused here on how superconducting qubits can be used for quantum information processing, there has of course also been tremendous activity in related fields, that briefly summarized below. One of these is {\bf quantum annealing}. Superconducting qubits also form the basis for certain quantum annealing platforms~\cite{Farhi2001}. Quantum annealing operates by finding the ground state of a given Hamiltonian (typically a classical Ising Hamiltonian), and this state will correspond to the solution of an optimization problem. By utilizing a flux-qubit type design, the company D-Wave have demonstrated quantum annealing processors~\cite{Johnson2011} which have now reached beyond 2000 qubits~\cite{DWaveSite}. The benchmarking of quantum annealers and attempts to demonstrate a quantum speedup for a general class of problems is a highly attractive field of research. \\
A parallel effort to the planar superconducting qubits has been the development of {\bf3D cavity-based superconducting qubits}. In these systems, quantum information is encoded in superpositions of coherent photonic modes of the cavity~\cite{Wang2016}. The cat states can be highly coherent due to the inherently high quality factors associated with 3D cavities~\cite{Pfaff2017}. This platform is characterized by a small hardware overhead for encoding a logical qubit~\cite{Heeres2017}, and it's kin to certain implementations of asymmetric error-correcting codes due to the fact that errors due to single-photon loss in the cavity is a tractable observable to decode. Using this architecture, several important advances were recently demonstrated including extending the lifetime of an error-corrected qubit~\cite{Ofek2016}.\\
Despite significant advancements in the improvement of qubit lifetimes and gate fidelity over the past decades, the need for error correction persists, especially as quantum processors move towards larger scales. Incorporating error correction into the quantum data necessitates the adoption of an {\bf error-correcting scheme}. Certain elements of the surface code quantum error correction framework have already been successfully demonstrated in the context of superconducting qubits, e.g. see Refs.~\cite{Riste2015,Takita2016}). However, a considerable challenge that remains is the demonstration of a logical qubit with a longer lifespan than its underlying physical qubits.
Despite the surface code's potential as a quantum error-correcting code due to its relatively robust fault tolerance threshold, it faces limitations in implementing a \emph{universal} gate set in a fault-tolerant manner. Consequently, error-corrected gates within the surface code need supplementation, such as incorporating a \textsf{T} gate through techniques like \emph{state distillation}~\cite{Bravyi2005}. Although gate-teleportation, a precursor to magic state distillation, has been demonstrated using planar superconducting qubits~\cite{Ryan2017}, showcasing distillation and injecting into a surface code logical state remains an unresolved challenge.\\
The continual evolution of new quantum codes represents a rapidly developing field. For more detailed insights, readers are encouraged to consult recent reviews, such as Ref.~\cite{Campbell2017}. Another pivotal step in the progression towards large-scale quantum processor architecture is the establishment of remote entanglement, facilitating the distribution of quantum information across different nodes within a quantum processing network~\cite{Axline2018}.\\
In the next year, one of the significant challenges facing superconducting qubits is the attainment of {\bf quantum computational supremacy}, as outlined by Preskill in 2012~\cite{Preskill2012}. The fundamental concept involves showcasing a computation utilizing qubits and algorithmic gates that surpasses the capabilities of classical computers, assuming certain plausible computational complexity conjectures. One possible review is the article by Harrow~\cite{Harrow2015}. A recent advancement toward achieving quantum supremacy was documented, utilizing 9 tunable transmons~\cite{Neill2018}.  The realization of quantum supremacy with a number of qubits ranging in number of hundreds would represent a remarkable achievement for the field of quantum computing.

\end{document}